\documentclass[12pt,preprint]{aastex}
\usepackage{graphicx}
\usepackage{emulateapj5}
\usepackage{apjfonts}
\usepackage{epsfig}
\usepackage{natbib}

\newcommand{\chisq}{\ensuremath{\chi^2}}

\def\gtrsim{\mathrel{\hbox{\rlap{\hbox{\lower4pt\hbox{$\sim$}}}\hbox{\raise2pt\hbox{$>$}}}}}

\newcommand{\halpha}{H\ensuremath{\alpha}}

\newcommand{\hbeta}{H\ensuremath{\beta}}

\newcommand{\hst}{\emph{HST}}

\newcommand{\kms}{km~s\ensuremath{^{-1}}}

\newcommand{\lya}{Ly \ensuremath{\alpha}}

\newcommand{\mbh}{\ensuremath{M_\mathrm{BH}}}

\newcommand{\msun}{\ensuremath{M_{\odot}}}

\newcommand{\nii}{[\ion{N}{2}]}

\newcommand{\oiii}{[\ion{O}{3}]}

\newcommand{\spitzer}{\emph{Spitzer}}

\def\lax{{$\mathrel{\hbox{\rlap{\hbox{\lower4pt\hbox{$\sim$}}}\hbox{$<$}}}$}}
\def\gax{{$\mathrel{\hbox{\rlap{\hbox{\lower4pt\hbox{$\sim$}}}\hbox{$>$}}}$}}

\slugcomment{Accepted version; April 11, 2014;
to be published in {\it The Astrophysical Journal}.}
\shorttitle{{\it NIR T2 QSOs}}
\shortauthors{Greene, ET AL.}

\begin{document}

\title{Near Infrared Spectra and Intrinsic Luminosities of Candidate Type II
Quasars at $2 < z < 3.4$}

\author{Jenny E. Greene\altaffilmark{1}, 
Rachael Alexandroff\altaffilmark{2}, Michael A. Strauss\altaffilmark{1},
Nadia L. Zakamska\altaffilmark{2}, Dustin Lang\altaffilmark{3}, 
Guilin Liu\altaffilmark{2}, Petchara Pattarakijwanich\altaffilmark{1},
Frederick Hamann\altaffilmark{4}, Nicholas P. Ross\altaffilmark{5}, 
Adam D. Myers\altaffilmark{6},
W. Niel Brandt\altaffilmark{7,9}, Donald York\altaffilmark{8}, 
Donald P. Schneider\altaffilmark{7,9}}

\altaffiltext{1}{Department of Astrophysical Sciences, Princeton University, 
Princeton, NJ 08544}
\altaffiltext{2}{Center for Astrophysical Sciences, Department of Physics and 
Astronomy, Johns Hopkins University, Baltimore, MD 21218, USA}
\altaffiltext{3}{McWilliams Center for Cosmology, Department of Physics, 
  Carnegie Mellon University, Pittsburgh, PA, USA}
\altaffiltext{4}{Department of Astronomy, University of Florida, Gainesville, 
FL 32611-2055, USA}
\altaffiltext{5}{Department of Physics, Drexel University, 3141 Chestnut Street, Philadelphia, PA 19104, USA}
\altaffiltext{6}{Department of Physics and Astronomy, University of Wyoming, Laramie, WY 82071, USA}
\altaffiltext{7}{Department of Astronomy and Astrophysics, The Pennsylvania
State University, 525 Davey Lab, University Park, PA 16802 USA}
\altaffiltext{8}{University of Chicago Astronomy and Astrophysics Department and Enrico Fermi Institute, 5640 So. Ellis Ave., Chicago, 60637}
\altaffiltext{9}{Institute for Gravitation and the Cosmos, The Pennsylvania 
State University, University Park, PA 16802}

\begin{abstract}
  We present $JHK$ near-infrared (NIR) spectroscopy of 25 candidate
  Type II quasars selected from the Sloan Digital Sky Survey, using
  Triplespec on the Apache Point Observatory 3.5m telescope, FIRE
  at the Magellan/Baade 6.5m telescope, and GNIRS on Gemini. At redshifts of $ 2 < z <
  3.4$, our NIR spectra probe the rest-frame optical region of these
  targets, which were initially selected to have strong lines of
  \ion{C}{4} and Ly~$\alpha$, with FWHM$<$2000~\kms\ from the SDSS pipeline. 
  We use the
  \oiii$~\lambda 5007$ line shape as a model for the narrow line region
  emission, and find that \halpha\ consistently requires a broad
  component with FWHMs ranging from 1000 to 7500~\kms.  Interestingly,
  the \ion{C}{4} lines also require broad bases, but with considerably
  narrower widths of $1000$ to $4500$~\kms. Estimating the extinction
  using the Balmer decrement and also the relationship in lower-$z$
  quasars between rest equivalent width and luminosity in the \oiii\ line, we find
  typical $A_V$ values of $0-2$ mag, which naturally explains the
  attenuated \ion{C}{4} lines relative to \halpha.  We propose that
  our targets are moderately obscured quasars.  We also describe one
  unusual object with three distinct velocity peaks in its \oiii\
  spectrum.
\end{abstract}

\section{Introduction}
\label{sec:Introduction}

A major stumbling block to understanding the accretion history of
supermassive black holes (BHs) over cosmic time is determining the
role of obscuration in the demographics of active galactic nuclei
(AGN). Optical surveys are quite successful at finding luminous blue
(unobscured Type I) quasars \citep[e.g.,][]{richardsetal2006,rossetal2013},
but are not sensitive to truly obscured (or Type II) quasars, where the quasar 
continuum and broad emission lines are completely hidden.
Sensitive 0.5-10 keV X-ray surveys can find modestly obscured
systems, but typically cover a limited solid angle, and thus are not
sensitive to rare luminous objects
\citep[e.g.,][]{brandthasinger2005,gillietal2007,xueetal2012,
  georgantopoulosetal2013}. Similar caveats apply to mid-infrared
(MIR) selection using \spitzer\
\citep[e.g.,][]{lacyetal2004,sternetal2005,donleyetal2008,
  donleyetal2012,lacyetal2013}, while \emph{WISE} can unambiguously select 
only the most luminous AGN 
\citep[e.g.,][]{sternetal2012,assefetal2013, yanetal2013}. Many X-ray--based studies
indicate a decreasing obscured fraction as a function of luminosity
over a broad redshift range
\citep[e.g.,][]{steffenetal2003,hasinger2008}, but at low redshift ($z
< 0.8$), luminous optically selected samples suggest comparable
numbers of obscured and unobscured systems \citep{reyesetal2008}. At
high redshifts, at the peak of quasar activity ($z \simeq 2-3$),
demographics are even more uncertain and a large-area survey of
luminous obscured quasar activity is required to determine the
obscured fraction.

Although unified theories state that Type~I and Type~II quasars
differ only in orientation
\citep[e.g.,][]{antonuccimiller1985,antonucci1993,urrypadovani1995},
there are many hints that Type II quasars may represent a special
phase in the growth of black holes.  There have been numerous suggestions,
both observational
\citep[e.g.,][]{sandersetal1988,canalizostockton2001,pageetal2001,ho2005,
stevensetal2005,veilleuxetal2009} and theoretical \citep[e.g.,][]{hopkinsetal2006},
that major galaxy mergers trigger an obscured phase of central BH
growth.  This obscured phase persists until the 
AGN grows powerful enough to expel all remaining gas out of
the surrounding galaxy, leading to an optically luminous quasar phase.
Samples of Type I quasars with moderate extinction hosted by
merging galaxies provide some support for this scenario
\citep[e.g.,][]{urrutiaetal2008}.  Again, larger
homogeneous samples of obscured quasars, at the peak epoch of BH
growth, are needed to statistically address the growth phase of these
objects.

Finally, some obscuration may occur in a torus near the AGN, while
some may be due to galaxy-scale dust \citep{rigbyetal2006}. The torus
geometry or porosity may also depend on luminosity
\citep[e.g.,][]{steffenetal2003,assefetal2013}. Larger samples of luminous
obscured quasars could address the full distribution of
obscuration as a function of bolometric luminosity.

To date, Type II samples at high redshift number in the tens,
including targets selected in the radio
\citep[e.g.,][]{mccarthy1993,urrypadovani1995,sternetal1999}, X-ray
\citep{normanetal2002,sternetal2002,bargeretal2003,iwasawaetal2012},
mid-infrared
\citep{lacyetal2004,sternetal2005,donleyetal2012,eisenhardtetal2012,assefetal2013,
  lacyetal2013} and optical
\citep{steideletal2002,bongiornoetal2010,hainlineetal2012,
  mignolietal2013}.  

\vbox{ 
\vskip +2mm
\hskip -3mm
\includegraphics[scale=.5,angle=0]{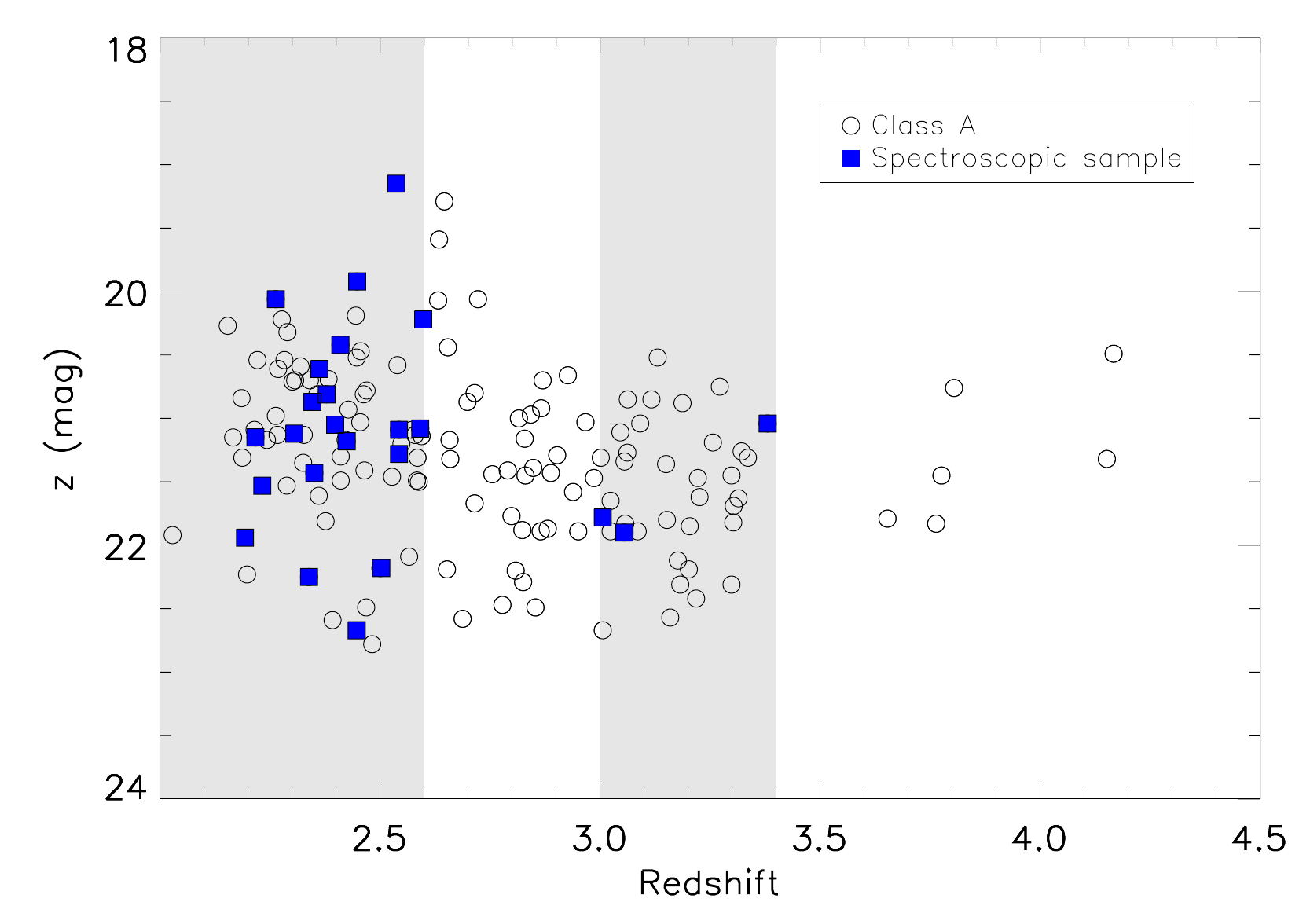}
\figcaption{
Redshift vs. $z-$band magnitude for the Alexandroff et al. Class A objects 
(open) and the sample spectroscopically targeted in the NIR (blue squares).  Shaded bands 
indicate redshift regions where strong rest-frame optical emission lines fall in 
atmospherically transparent regions of the spectrum.  Our spectroscopic 
subsample is representative of the full distribution.
\label{fig:redshiftmag}}
}
\vskip 5mm
\noindent
The operational definition of obscured quasar
depends on the selection technique.  The typical X-ray definition of
\ion{H}{1} column density $N_H > 10^{22}$ cm$^{-2}$
\citep[e.g.,][]{uedaetal2003} apparently corresponds to an $A_V
\approx 0.47$ in AGN \citep{maiolinoetal2001,assefetal2013}.  In
contrast, optically selected Type II Seyfert galaxies typically have
$A_V > 10$
\citep[e.g.,][]{veilleuxetal1997,zakamskaetal2005}. Finally,
near-infrared selection criteria yield moderately reddened quasars
with broad \halpha\ and typical $A_V \sim 0.3-6$
\citep[e.g.,][]{banerjietal2012, glikmanetal2012}.

In this paper, we focus on Type II candidates selected in the
rest-frame UV, to explore the $A_V$ distribution of objects selected
on the basis of narrow UV emission lines. Low-redshift ($z < 0.8$)
Type II quasars were successfully discovered in large numbers in the
Sloan Digital Sky Survey \citep[SDSS; ][]{yorketal2000} based on their
strong and narrow \oiii$~\lambda 5007$\AA\ emission
\citep{zakamskaetal2003,reyesetal2008}, and subsequently shown to be
bona fide obscured quasars
\citep{zakamskaetal2005,zakamskaetal2006,zakamska08}. Not until the
Baryon Oscillation Spectroscopic Survey
\citep[BOSS;][]{eisensteinetal2011,dawsonetal2013}, which
spectroscopically targeted quasars down to a magnitude limit of $g <
22$ or $r < 21.85$ \citep{rossetal2012}, did it become possible to
select Type II quasar candidates with $z > 1$ with the SDSS.  In
\citet[][hereafter Paper I]{alexandroffetal2013} and this work, high
redshift Type II quasar candidates are identified based on the
presence of strong and narrow high-ionization lines in their
rest-frame UV spectra (e.g., \ion{C}{4}).

In Paper I, we presented a sample of 145 Type~II quasar candidates
selected from the BOSS survey based on their narrow ($< 2000$~\kms)
\ion{C}{4} and Ly $\alpha$ emission.  The narrow linewidths and
high rest equivalent widths (EWs) of the sample objects bear strong
resemblance to those of other samples of Type II quasars. Furthermore,
in the two objects we observed with a spectropolarimeter, we detected
continuum polarization of $\sim 3\%$, inconsistent with typical
unobscured quasars. On the other hand, our BOSS Type II quasar
candidates are too luminous and blue in the UV continuum to be
explained by galaxy light alone.  They have typical rest-frame UV
continuum luminosities of $-24$ AB mag at 1450~\AA, as compared to
magnitudes of $\sim -22.5$ AB mag \citep{okegunn1983} for the most
luminous UV-selected galaxies at similar redshifts
\citep{shapley2011}.  The AGN must contribute some UV light, whether
it be directly transmitted or due to scattered light. Furthermore, the
UV line ratios are more akin to Type I rather than Type II objects.
Finally, in one object with broad spectral energy distribution (SED)
coverage, the optical/UV is weaker than seen in typical Type I objects
indicating some obscuration, but more prominent than in typical Type
II objects.  In the same sense, the optical/NIR colors of our Type II
candidates are similar to unobscured quasars.  Thus, based on the
rest-frame UV spectra alone, it is difficult to determine the true
nature of these sources.

\begin{figure*}
\vbox{ 
\vskip -8mm
\hskip +25mm
\includegraphics[scale=.8,angle=0]{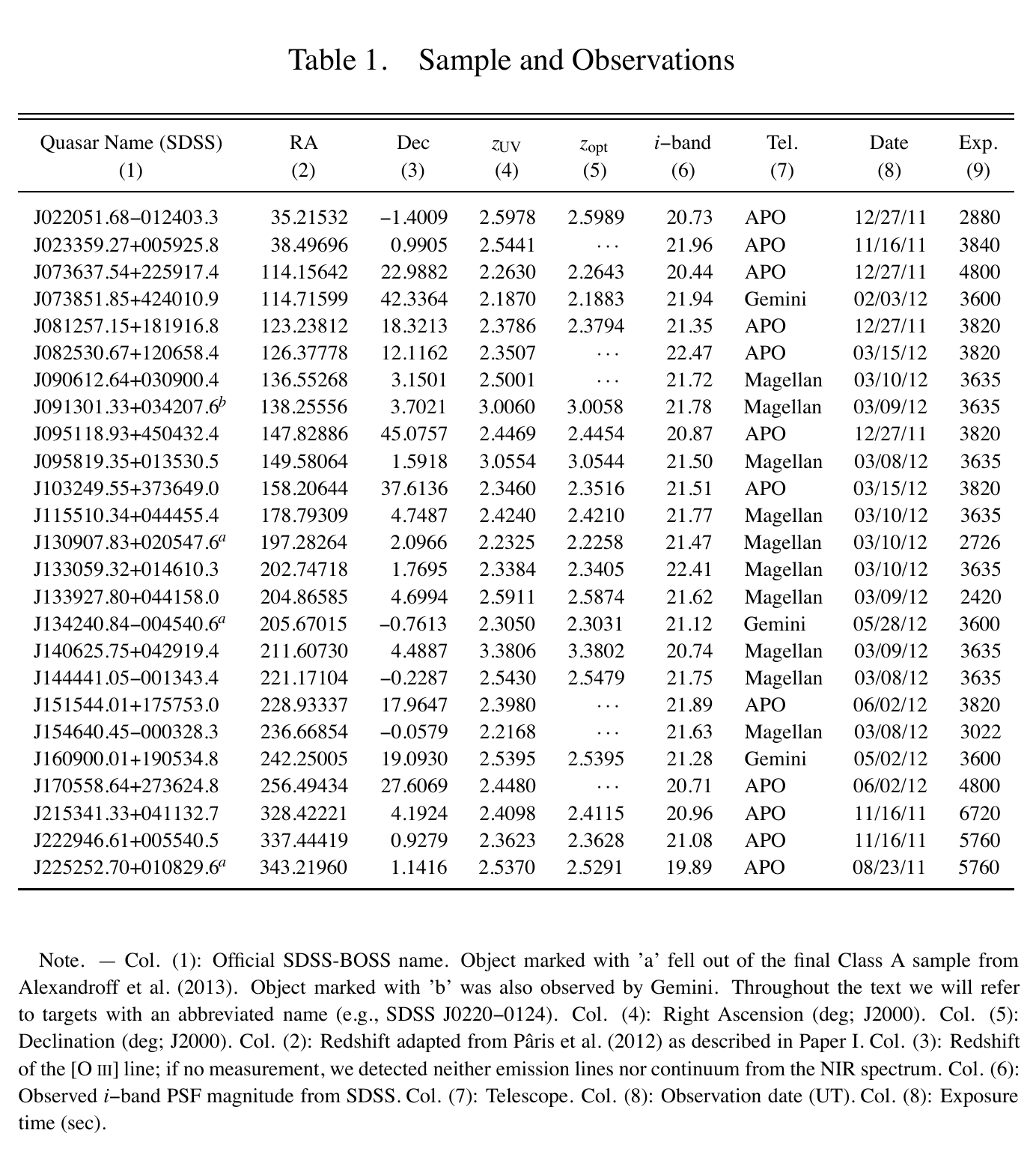}
}
\vskip -0mm
\label{tab:sample}
\end{figure*}

Here, we present NIR spectroscopy that probes the rest-frame optical
spectra of 25 of the Type~II quasar candidates presented in Paper I.
We use three NIR echellettes, Triplespec \citep{tspec2004} on the 3.5m
at Apache Point Observatory (APO), the Folded-port InfraRed Echellette
\citep[FIRE;][]{simcoeetal2013} at Magellan, and Gemini Near Infrared 
Spectrograph \citep[GNIRS;][]{eliasetal2006} on Gemini North, all
three of which afford us $JHK$ spectroscopy in a single observation. We
simultaneously measure \hbeta, \halpha, and the strong and ubiquitous
\oiii$~\lambda 5007$ line in the majority of our targets.  The \oiii\
line luminosity is known to correlate with the intrinsic luminosity of
the quasar \citep{yee1980,heckmanetal2004,liuetal2009}, while the
strength and width of the Balmer lines provide new insight into the
level and scale of the extinction.  Finally, the \oiii\ line shape
unambiguously traces the low-density (narrow-line region) gas
kinematics, and thus allows us to characterize any additional
(broader) components in the permitted lines.  In fact, we specifically
targeted two galaxies with multiple velocity peaks in the \ion{C}{4}
and/or \lya\ line (Paper I), to determine whether the peaks are caused
by absorption or real kinematic structure in the gas.

\begin{figure*}
\vbox{ 
\vskip 0mm
\hskip +10mm
\includegraphics[scale=.75,angle=90]{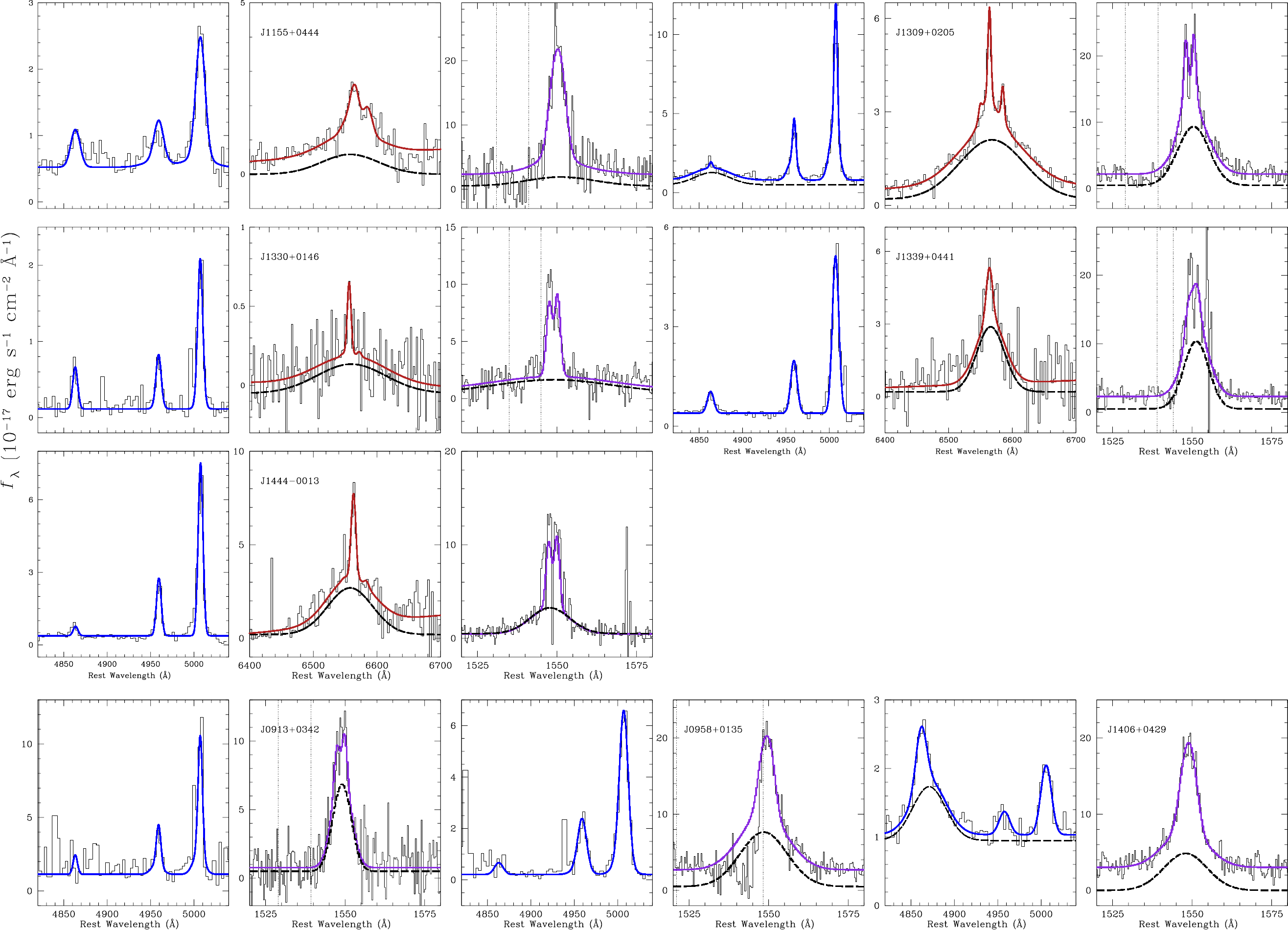}
}
\vskip -0mm
\figcaption[]{The top three rows show 
fits to the [O {\tiny III}], \halpha, and C~{\tiny IV} regions of all five of the Magellan 
spectra with \halpha\ in the spectral region, while the C~{\tiny IV} is 
from the SDSS spectra.  In the bottom row, we present the three 
higher redshift targets for which \halpha\ was not observed. The 
original spectra are binned by three pixels (thin black solid lines), and
our best overall model fit (thick blue, red, or magenta solid lines for 
[O {\tiny III}], \halpha, and C~{\tiny IV} respectively), and when present the broad 
components are shown offset for clarity (thick dotted lines). 
In the fits to the \halpha\ and C~{\tiny IV} regions, 
the narrow component is constrained to match the [O {\tiny III}] model shape.
The dashed vertical lines indicated the regions that 
  were masked in fitting C~{\tiny IV}. See \S \ref{sec:Fitting} for details.  
\label{fig:nirfits}}
\end{figure*}

The paper proceeds as follows.  In \S \ref{sec:Observations} we
present properties of the sample and details of the observations. In
\S \ref{sec:Reductions} we discuss the data reduction, while in \S
\ref{sec:Fitting} we outline our line-fitting technique. We present
our results in \S \ref{sec:Results}, and discuss the implications for obscured 
quasars in \S \ref{sec:Discussion}. We assume a concordance 
cosmology with H$_0 = 70$~\kms~Mpc$^{-1}$, $\Omega_{\rm M} = 0.3$, and 
$\Omega_{\rm M} = 0.7$ \citep{dunkleyetal2009}.

\begin{figure*}
\vbox{ 
\vskip 0mm
\hskip +15mm
\includegraphics[scale=.75,angle=90]{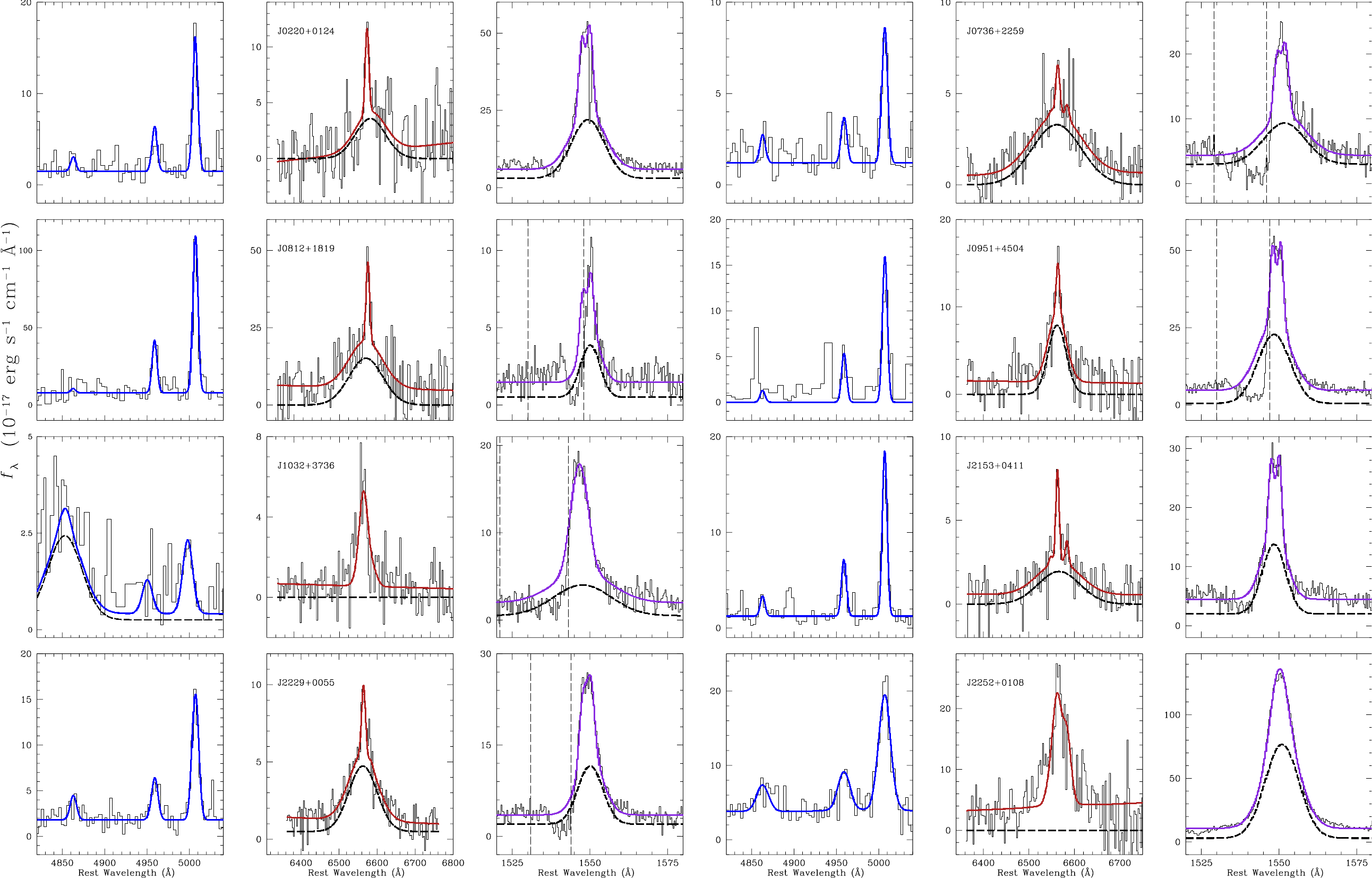}
}
\vskip -0mm
\figcaption[]{
Same as Figure \ref{fig:nirfits} above, but for twelve galaxies
  observed with APO/Triplespec.  We show the original spectra in units
  of $10^{-17}$~erg~s$^{-1}$~cm$^{-2}$~\AA$^{-1}$ (thin black solid
  lines), our best overall model fit (thick blue solid lines), and
  when present the broad components are shown offset for clarity
  (thick dotted lines).  In the case of the \halpha\ and C~{\tiny IV}
  regions, the narrow component is constrained to match the [O {\tiny
    III}] model shape. The dashed vertical lines indicated the regions that 
  were masked in fitting C~{\tiny IV}. See \S \ref{sec:Fitting} for details.  
\label{fig:nirfitsAPO}}
\end{figure*}

\section{Sample Description and Observations}
\label{sec:Observations}

Our parent Type II sample is discussed in detail in
\citet{alexandroffetal2013}; we summarize our selection criteria
briefly here. We started with SDSS/BOSS spectra
\citep{gunnetal2006,smeeetal2013}, first released in the 9th Data
Release \citep[DR9;][]{ahnetal2012DR9,parisetal2012}, and selected all
the SDSS/BOSS objects in DR9 with both \lya\ and \ion{C}{4} in the
spectra (corresponding to $z > 2.0$), a reliable spectroscopic
pipeline fit \citep[i.e., the flag ZWARNING = 0;][]{boltonetal2012} and $5
\sigma$ detections in both \lya\ and \ion{C}{4}.  We selected the
targets with FWHM (\ion{C}{4} and \lya) $< 2000$~\kms; while somewhat
arbitrary, this cut is designed based on observations of luminous Type
II quasars at $z \approx 0.5$
\citep{greeneetal2011,liuetal2013a,liuetal2013b}.  The BOSS spectrum
of each candidate was then visually inspected. Objects with extreme
absorption features are classified as broad absorption line (BAL)
quasars (14\% of the narrow-line objects), while those with
strong permitted Fe emission and strong blue continua were classified
as narrow-line Seyfert 1 galaxies \citep[NLS1s; e.g.,][56\% of
  the narrow-line objects]{osterbrockpogge1985}.  After removing
these two classes, the resulting sample comprises 452 objects that are
strong or possible cases for Type II quasars.

Even after removing the most obvious BALs and NLS1s, some objects
still appeared ambiguous (for instance, the \ion{C}{4} line had a
broad base). Those most likely to be Type II quasars were classified
as Class A (145 objects), while the ambiguous sources were classified
as Class B (307 objects).  As shown in Paper I, this visual
classification separates the sources by linewidth and continuum
strength, although ultimately it is somewhat subjective. The median
FWHM of the CIV line among the Class A candidates is $\langle {\rm FWHM}
\rangle =1260$~\kms. The mean redshift of these two samples is
$\langle z \rangle =2.70$, with a redshift range from 2.03 to
4.23. Almost all the objects in both Class A and Class B were selected
for spectroscopy as quasar candidates, using the algorithms described
in \citet{rossetal2012}, with only four Class A candidates selected under 
various auxiliary programs (e.g., a color-based high-redshift quasar search).

All but three of the objects presented here are Class A targets.  There are
three exceptions that do not appear in Paper I, but were selected in an
early version of this work using less stringent \ion{C}{4} and Lyman
$\alpha$ restrictions.  In total, we have obtained NIR spectra for 12
objects with APO, 10 with Magellan, and three additional objects 
(plus one repeated object) with Gemini (Table 1).
The final 25 targets span the luminosity and redshift range of the
full sample.  Type~II quasar candidates were selected for NIR
spectroscopic observation based on a number of criteria, and the
sample is summarized in Figure \ref{fig:redshiftmag}.  Since many
Type~II quasars in the literature have been found via radio 
selection, we focused on the
majority of targets that are undetected in FIRST 
\citep[comprising 98\% of the class A sample;][]{beckeretal1995}, 
although given the limited FIRST sensitivity and
the faint UV continua, we cannot rule out that many of the non-radio
detections are actually radio loud.  We
then selected targets in the redshift ranges $2 < z < 2.6$ or $3.0 < z
< 3.4$, to ensure that the emission lines of interest, H$\beta$,
[OIII] and H$\alpha$ (at low redshift), fell into windows of high
atmospheric transparency. Within these wavelength windows, we further selected
windows where none of the strong emission lines coincide with strong
sky emission lines.  Finally, we targeted two interesting targets with
multiple velocity components in both the Lya and \ion{C}{4} lines (\S
\ref{sec:Multi}). From the UV spectra alone, we suggested that the
observed velocity structure of the CIV line was likely due to
superposed absorption features (Paper I). Another possibility was that
the multiple peaks corresponded to multiple physical components,
perhaps in the process of forming a new galaxy. Our NIR spectra allow
a test of these hypotheses.

For the Triplespec targets, we additionally imposed a $z-$band
magnitude limit of $z < 21$, such that we could identify targets in
the guider (although occasionally blind offsets were used to place the
target in the slit).  For the FIRE data, we prioritized
fainter targets, down to $z \approx 22.7$ mag. In cases where
multiple targets satisfied these criteria, priority was given to those
with matches in various X-ray surveys, those in our approved \hst\
imaging program, and those with the multi-peaked emission lines
described above. Finally, we prioritized those few targets with nearby
tip-tilt stars for future adaptive-optics observation.  Below we
describe the observations taken with each instrument individually.

\vbox{ 
\vskip 0mm
\hskip -1mm
\includegraphics[scale=.55,angle=0]{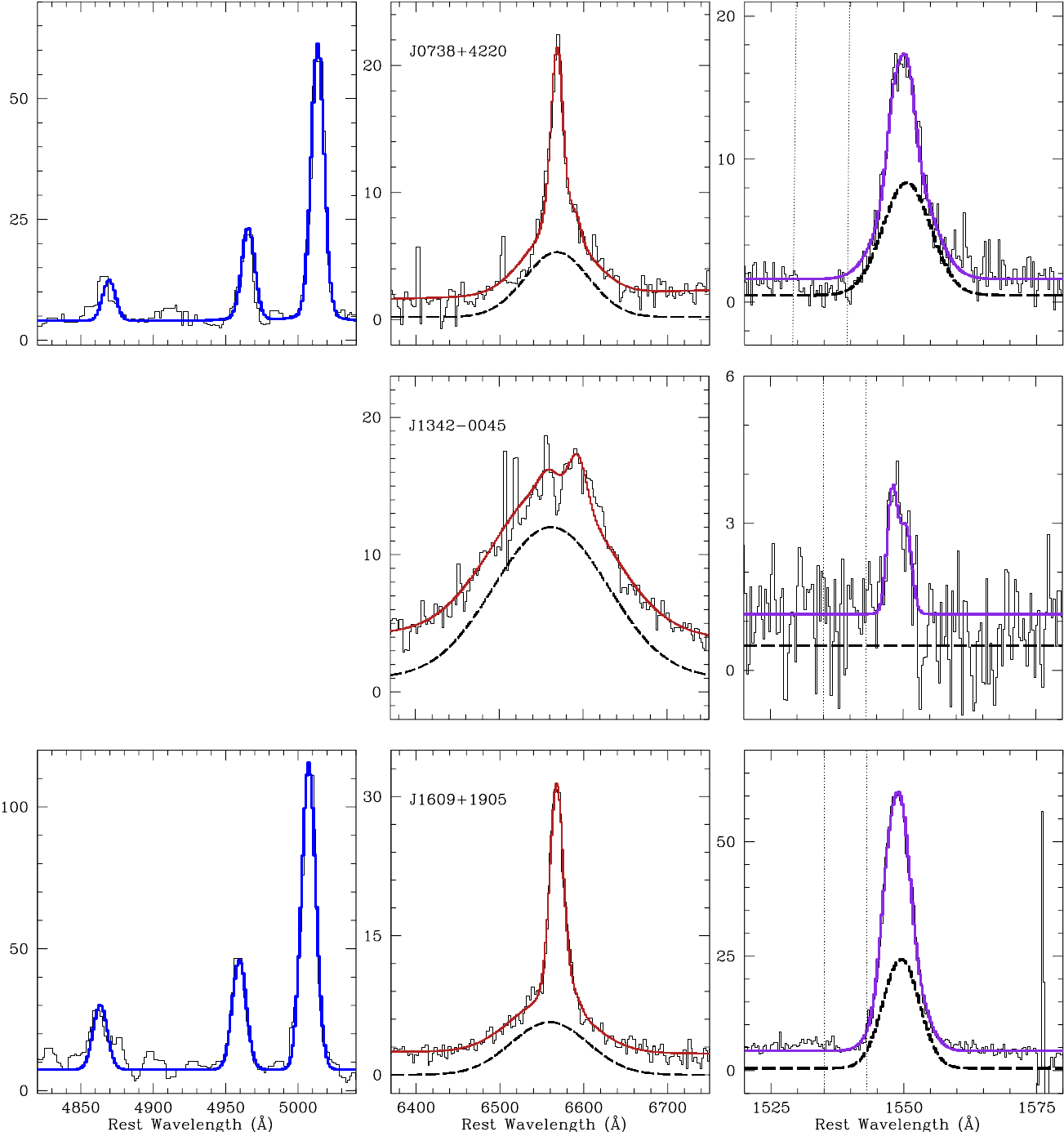}
}
\vskip -0mm
\figcaption{As in Figure \ref{fig:nirfits} and Figure \ref{fig:nirfitsAPO} 
above we show the H$\beta$ (left) and H$\alpha$ (center)
regions of the spectra taken with Gemini and the C{\tiny IV} region (right) 
as observed by the SDSS. The y-axis scalings are arbitrary since we 
have not flux calibrated the Gemini data. We 
display the unbinned data (black histogram), the models (in blue, red, and 
magenta respectively), and any broad component to the H$\alpha$ and 
C{\tiny IV} (dashed).  In the case of J1342$-0045$, no narrow [O{\tiny III}],
H$\beta$, or H$\alpha$ is detected and even the C{\tiny IV} is weak, but 
broad \halpha\ is quite prominent.
\label{fig:nirfitsGEM}
}
\vskip 5mm

\subsection{Triplespec}

Triplespec is a cross-dispersed NIR spectrograph \citep{tspec2004}
mounted on the Apache Point 3.5m telescope in New Mexico.  It observes
simultaneously from 0.95~$\mu$m to 2.46~$\mu$m (separated into five
spectral orders) with a spectral resolution of $R\approx5000$.  We
used a $1\farcs1$ or $1\farcs5$ slit depending on observing
conditions.  To maximize the signal-to-noise ratio for our relatively
long exposure times, we used Fowler=8 sampling.

We targeted 12 of our candidates using the Triplespec instrument.  The
data were taken over the course of four nights: August 23, November
15, and December 26 2011 and March 14 2012 (Table 1).  Of the 12
objects that we targeted for spectroscopy, we succeeded in detecting
line and/or continuum emission for eight. In the following discussion,
we will focus only on the detected targets.  However, we do note 
that we preferentially failed to detect fainter targets (with $i \gtrsim 22$ mag). 
On the other hand, we have succeeded in detecting sources at 
these faint magnitudes with Magellan and Gemini.

We observed with a standard ABBA sequence to facilitate sky
subtraction, with individual exposure times of 240 sec. We observed
each target over the course of 1-2 hours, for total on-source exposure
times of 1-1.5 hours (Table 1).  We bracketed each target with
observations of a telluric standard, typically an A0V star with
magnitudes of $V \approx 7$ to 9 mag, at a similar airmass
\citep[e.g.,][]{cushingetal2004}.  This star is used both to remove
telluric absorption and as a flux calibrator.

\subsection{Folded-port Infrared Echellette}

The Folded-port Infrared Echellette \citep[FIRE;][]{simcoeetal2013} is
a NIR spectrograph used on the 6.5m Baade-Magellan Telescope.  In
high-resolution mode, it simultaneously observes from 0.8$\mu$m to
2.5$\mu$m over 21 spectral orders with a spectral resolution of
$50$~\kms\ or $R \approx 6000$ for the $0\farcs75$ slit that we used.
Typical exposure times were 900 sec.  In this case we adopted the
Sample Up The Ramp (SUTR) observing technique. SUTR samples the array
continuously and the signal is derived as a fit to the accumulated
charge over the elapsed time, reducing the read noise significantly.
SUTR also has smaller overheads than Fowler Sampling. Our FIRE data
were obtained over three half nights on March 7-9 2012. The conditions
were good to excellent, with typical seeing of $0\farcs6$ in the
optical.  As above, we used a standard ABBA observing sequence and
bracketed our science exposures with A0V standard stars, and we
typically spent 1 hour of on-source time per target.

\subsection{Gemini Near-infrared Spectrograph}

We obtained near-infrared spectra of four objects with the Gemini
Near-Infrared Spectrograph \citep[GNIRS;][]{eliasetal2006} on the Gemini
North 8.1 meter telescope on the summit of Mauna Kea, Hawaii.  We used
the 32 l/mm grating with a cross-dispersion and a $0\farcs45$ slit, giving a
resolution of $R \approx 1200$, and covering the full wavelength range of the $J$,
$H$, and $K$ bands.  The objects were chosen with redshifts such that
H$\alpha$ and the H$\beta$-\oiii\ complex fell in regions of high
atmospheric transparency.  Each object was observed in queue mode for
a total of 60 minutes in a series of nodded exposures along the slit
for 300 seconds each, under non-photometric conditions of moderate
seeing.

\section{Reductions}
\label{sec:Reductions}

The data reduction utilized pipelines described in
\citet{vaccaetal2003} and \cite{cushingetal2004} for Triplespec and
\citet{simcoeetal2012} for FIRE. The Gemini spectra were reduced and
calibrated using the Gemini package for IRAF, following the standard
example script for GNIRS long-slit observations described in the task
{\it gnirsexamples}.  We describe the overall procedure utilized for
all data sets, highlighting any differences. In brief, dome flats were
obtained and combined in a standard manner, including dark
subtraction, to perform a flat-field correction. Wavelength
calibration was based on the airglow $OH$ lines from the Earth's
atmosphere; wavelength solutions are derived for each target
individually.  The FIRE pipeline performs a 2D fit to the spatial
curvature following \citet{kelson2003}.

We extract our targets as unresolved point sources. AB pairs are
differenced to remove airglow emission lines, and then
each spectrum is extracted.  Extraction positions are determined for
each exposure due to occasional drift of the object position from
exposure to exposure in the Triplespec data, and then all AB pairs are
summed (or median-combined in the case of the Gemini 
spectra). Telluric correction proceeds in an identical manner in the
Triplespec and FIRE pipelines \citep{vaccaetal2003}.  We create a telluric correction
spectrum by smoothing and scaling the spectrum of Vega to match the
observed standard star; dividing by this model leaves a telluric
correction spectrum. The model is shifted in velocity and then the
target spectrum is divided by this matched telluric spectrum. Flux
calibration based on published photometry of the star is performed 
simultaneously.  All orders are merged and bad pixels are
removed.  Finally, the spectra are binned onto a uniform wavelength
grid. The Gemini data are not flux calibrated.

The nominal $K-$ band magnitudes, as estimated from our spectra, range
from $K_{\rm AB} \approx 21.5$ to $18.8$ mag.  In order to verify the
flux calibration of our near-infrared spectra from APO and Magellan we
matched our sample with the UKIRT Infrared Deep Sky Survey
\citep[UKIDSS;][]{lawrenceetal2007}. 

\vbox{ 
\vskip +8mm
\hskip 0mm
\includegraphics[scale=.45,angle=0]{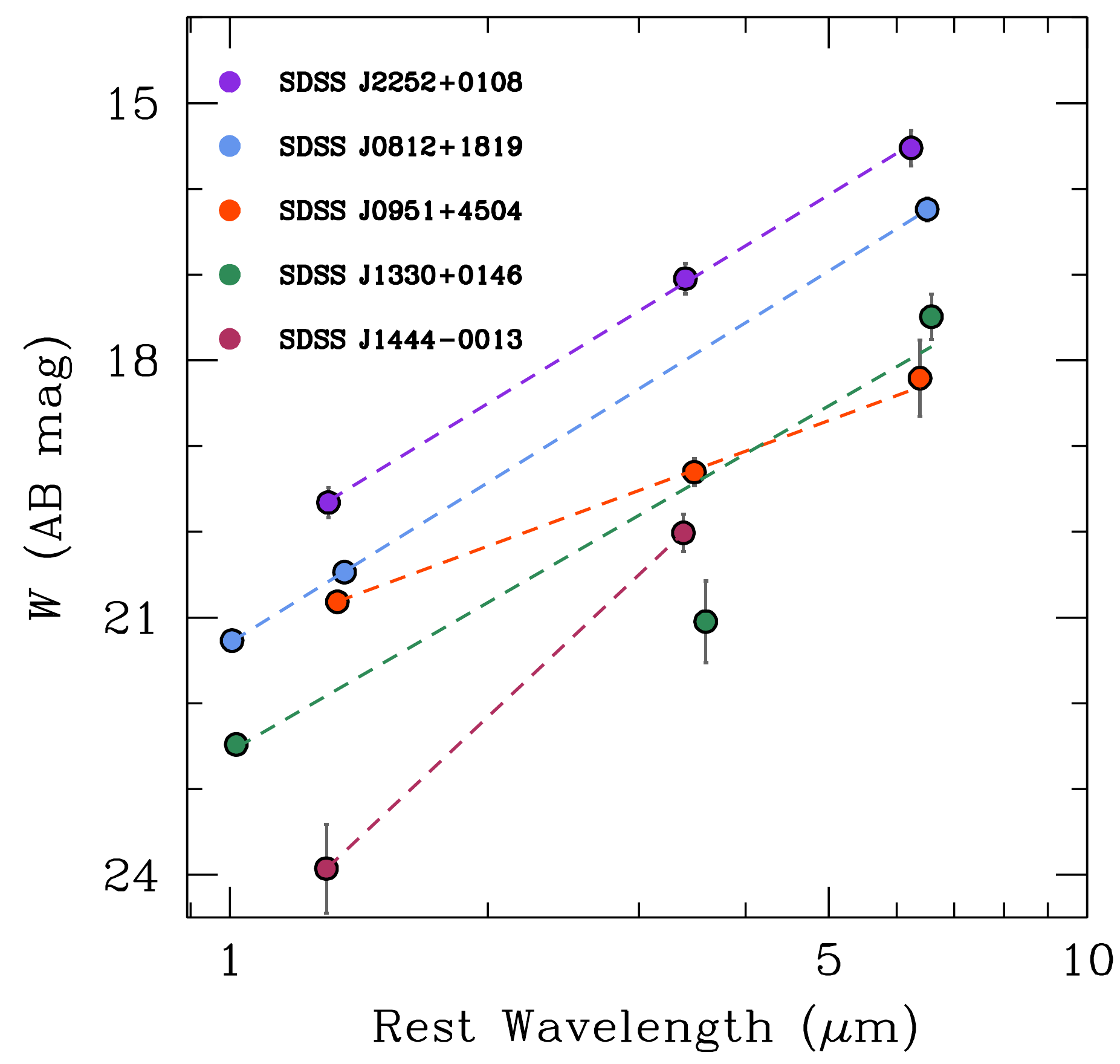}
}
\vskip -0mm
\figcaption[]{
Example power-law fits to \emph{WISE} photometry (AB mag).  Each
  source is shown in a different color, with dotted lines showing our
  fits.  Only three bands are fitted because we exclude those 
  with rest wavelengths $< 1 \micron$. 
  We use these power-law fits to find the luminosity at a rest
  wavelength of 3$\micron$. 
  These objects span the range of \emph{WISE} photometry quality in the sample.
\label{fig:wisefit}}
\vskip 5mm
\noindent
There are four matches, one
observed with APO, and the other three with Magellan.  We used the
filter transmission data from \citet{hewettetal2006} to determine
integrated fluxes from our APO and Magellan spectra.  The APO target
SDSS J222946.61+005540.5 (from now on we will abbreviate the SDSS
names as SDSS J2229+0055) has a spectrophotometric $H-$band magnitude
that agrees with the UKIDSS measurement to within 0.2 mag. Three
objects observed with Magellan had matches in the UKIDSS survey. One
(SDSS J1406+0429) has $H$ and $K-$band magnitudes that agree within
0.2 magnitudes with the UKIDSS survey.

The analysis of the other two objects is more complicated.  SDSS
J0958+0135 was observed with COSMOS \citep{scovilleetal2007} and so
has additional data (Paper I).  The UKIDSS magnitude of $H=20.2$ mag
(AB) agrees with the UKIRT value of $H=20.0$ mag quoted in Paper I
\citep{ilbertetal2009}, but is two magnitudes brighter than our
$H=21.97$ mag. Such a large discrepancy is difficult to attribute to
slit losses alone. We do detect line emission from this object, so it
was at least partially in the slit. The apparent SED is
quite red; it has SDSS $z = 21.9$ mag. In principle it is possible
that the strange colors are caused by variability, and indeed our sources
show striking similarities with the UV spectrum of NGC 5548 in the low
state \citep{goadkoratkar1998}.  On the other hand, the UKIDSS
and UKIRT photometry agree and were taken at different epochs.

\begin{figure*}
\vbox{ 
\vskip +1mm
\hskip +5mm
\includegraphics[scale=.75,angle=0]{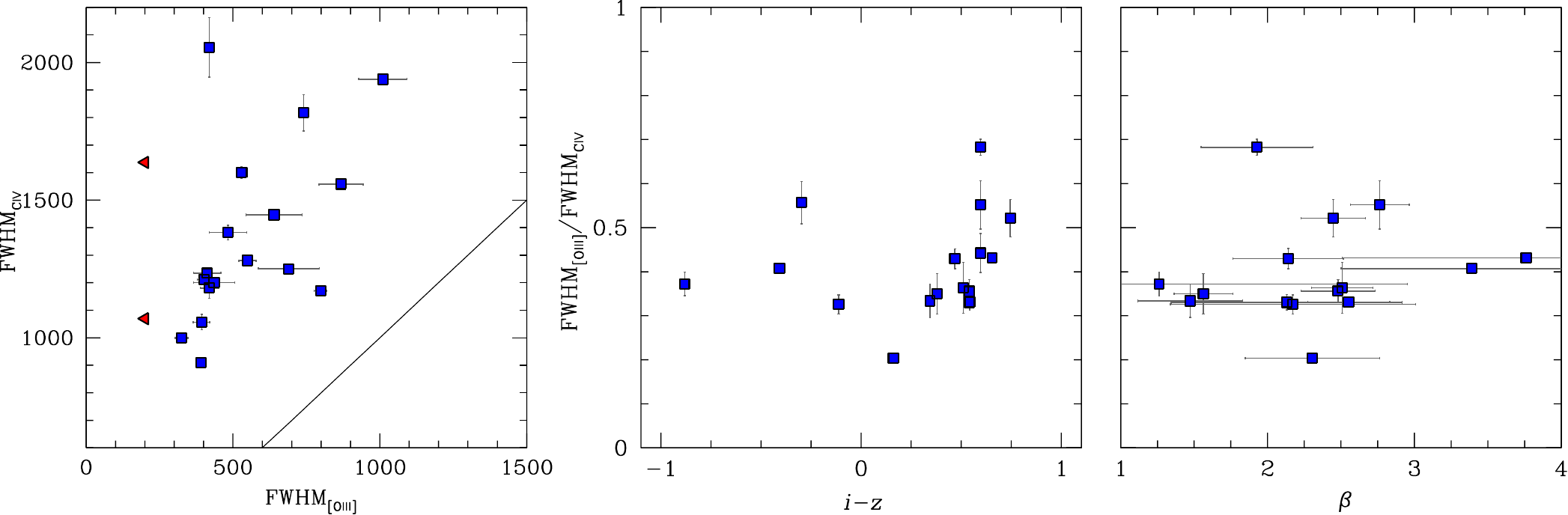}
}
\vskip -0mm
\figcaption[]{
{\it Left}: Comparison of FWHM$_{\rm [O {\tiny III}]}$ and 
FWHM$_{\rm C {\tiny IV}}$, as measured from the full profile fit to the 
broad and narrow components of C{\tiny IV}. The upper limits on [O {\tiny III}] 
(red triangles; SDSS J1032+3736 and SDSS J1342$-$0045) are 
simply assigned a FWHM of 200 km~s$^{-1}$ 
for illustrative purposes. The solid line demonstrates the 1:1 relation.
{\it Middle and Right}: 
Ratio of FWHM$_{\rm C {\tiny IV}}$ to FWHM$_{\rm [O {\tiny III}]}$ as a function of 
rest-frame UV color ($i-z$) and the MIR spectral slope $\beta$ that we fit to 
the four \emph{WISE} bands (\S \ref{sec:wise}). Both SDSS J1406+0429 and 
SDSS J0906+0309 are missing from the right-hand panel, since they 
do not have sufficient \emph{WISE} detections to measure a slope.
We see no correlation between continuum color and the dominance of the 
broad line. 
\label{fig:fwcolor}}
\end{figure*}
\vskip 5mm

Our final match, SDSS J1339+0441, observed with Magellan, is also
considerably fainter in the spectral continuum than the imaging, with
UKIDSS finding $H=20.6$ mag as opposed to $H=21.4$ mag from our FIRE
spectrum.  Again, the source is quite faint in the observed red, with
SDSS $z=21.1$ mag, but apparently brightens by a magnitude between the
$z$ and $H$ bands. With so few points of comparison, it is difficult
to diagnose the origin of these discrepancies.  We are in the process
of obtaining NIR imaging for a larger sample. In the meantime, 
our flux calibration, particularly for the FIRE data, currently
has at least a factor of two uncertainty.

\section{Line Fitting and Flux Measurements}
\label{sec:Fitting}

To obtain continuum flux and luminosity as well as emission line
widths, fluxes and luminosities, we fit Gaussian+continuum models to
the \oiii, \halpha, and \nii\ lines (Table 2). The \oiii$~\lambda
\lambda 4959, \, 5007$ lines have low critical density, and thus
unambiguously trace the kinematics in the low-density narrow-line
region.  We accordingly fit the \oiii\ lines first, fixing their
relative wavelengths to laboratory values and their line ratio to 1:3.
We allow up to two Gaussian components for this fit
\citep[following][]{greeneho2005gas}, and simultaneously model the
\hbeta\ line with the same shape. We minimize \chisq\ using
Levenberg-Marquardt minimization as implemented by MPFIT
\citep{markwardt2009}.  In most cases the \hbeta\ line is
too weak to allow us to model a broad component.  The exceptions are
SDSS J1309+0205, where we detect a strong \halpha\ line and a broad
component to \hbeta, SDSS J1406+0429, where the \hbeta\ line is
stronger than the \oiii\ line, and SDSS J1032+3736, where \oiii\ is
not significantly detected.  In these cases, \hbeta\ is allowed an additional 
broad Gaussian component, whose width is determined independently 
of \halpha.

We derive uncertainties on all fit parameters using Monte Carlo
simulations.  We start with the best-fit model, and then create 500
artificial spectra by adding Gaussian random noise generated from the
error spectrum.  We fit each of these artificial spectra in an
identical fashion, and the errors we quote enclose 68\% of the
artificially generated values.  Errors on all parameters are derived
in this manner.

We adopt the \oiii\ fits as a model for the narrow components of
\halpha+\nii.  Again the relative wavelengths are fixed, as is a ratio
of 1:3 for the \nii$~\lambda 6548$/\nii$~\lambda 6584$ lines.  Only the
overall redshift and the flux scale of the narrow \halpha\ and \nii\
lines, along with the local continuum, are free parameters.  In the
two cases where \oiii\ is undetected, we model the narrow-line region
as a single Gaussian with $\sigma$ allowed to vary. We then perform a
second fit including a broad \halpha\ component as well, represented
as a Gaussian with the same redshift to within 250~\kms. It is always
challenging to determine whether or not a given line includes a broad
component \citep[e.g.,][]{hoetal1997,zakamskaetal2003,greeneho2004,
  haoetal2005,reinesetal2013}.  As above, we run Monte Carlo
simulations to determine the errors in each parameter.  We deem the
broad component significant if $\chi^2_{\rm no broad} - \chi^2_{\rm
  broad} > 4$ and the broad-line flux is detected at $> 3 \sigma$
confidence.  Note that we add only two free parameters, so this
minimum increase in $\chi^2$ should represent a significant
improvement in the fit.  In practice, the improvement is far larger
than this in nearly all cases. Applying these two criteria, all but
two of the 16 quasars that include \halpha\ in their bandpass require
a separate broad \halpha\ component. The two exceptions are SDSS
J1032+3735 and SDSS J2252+0106.  In the case of SDSS J1032+3735, we do
not detect the \oiii\ line significantly, so the decomposition is
suspect.  Neither object stands out in other regards.  The line fits
to the Magellan spectra are shown in Figure \ref{fig:nirfits}, to
the APO spectra in Figure \ref{fig:nirfitsAPO}, and to the Gemini spectra 
in Figure \ref{fig:nirfitsGEM}.

\subsection{C{\small IV} Fits}

We then fit the \ion{C}{4}$~\lambda \lambda 1548, 1551$\AA\ lines from
the SDSS spectra.  We model the \ion{C}{4} line as a doublet, with a
velocity splitting of $\sim 500$~\kms\ fixed to the laboratory values,
and a line ratio of 0.9:1 (red:blue) taken from NIST
\citep{NIST_ASD}\footnote{http://physics.nist.gov}.  While this line ratio
is most appropriate for the high densities and/or high line optical
depths of the broad-line region, we do not have adequate S/N to 
independently constrain the narrow-line ratios \citep{hamannetal1995}. 
Each component is modeled with the same line width, so we
introduce no additional free parameters.  As above, we assume that the \oiii\
line traces the kinematics of the narrow-line region, and constrain
any narrow component of \ion{C}{4} to have that shape.  We
simultaneously allow the code to fit the residual flux with a single
Gaussian for each member of the doublet, with relative wavelengths and
line ratios fixed but all other components left as free parameters,
only constraining the broad-line central velocity to fall within $\pm
250$~\kms\ of the narrow line.  A serious complication arises from the
prevalence of absorption on the blue side of the \ion{C}{4} line.  We
mask the absorption regions by hand, based on the asymmetry in the
line profiles.  The masked regions are indicated as dashed lines in
Figures \ref{fig:nirfits} and \ref{fig:nirfitsAPO}.  Errors are derived
via fits to simulated spectra as for \halpha\ and \oiii.

\begin{figure*}
\vbox{ 
\vskip +1mm
\hskip +5mm
\includegraphics[scale=.75,angle=0]{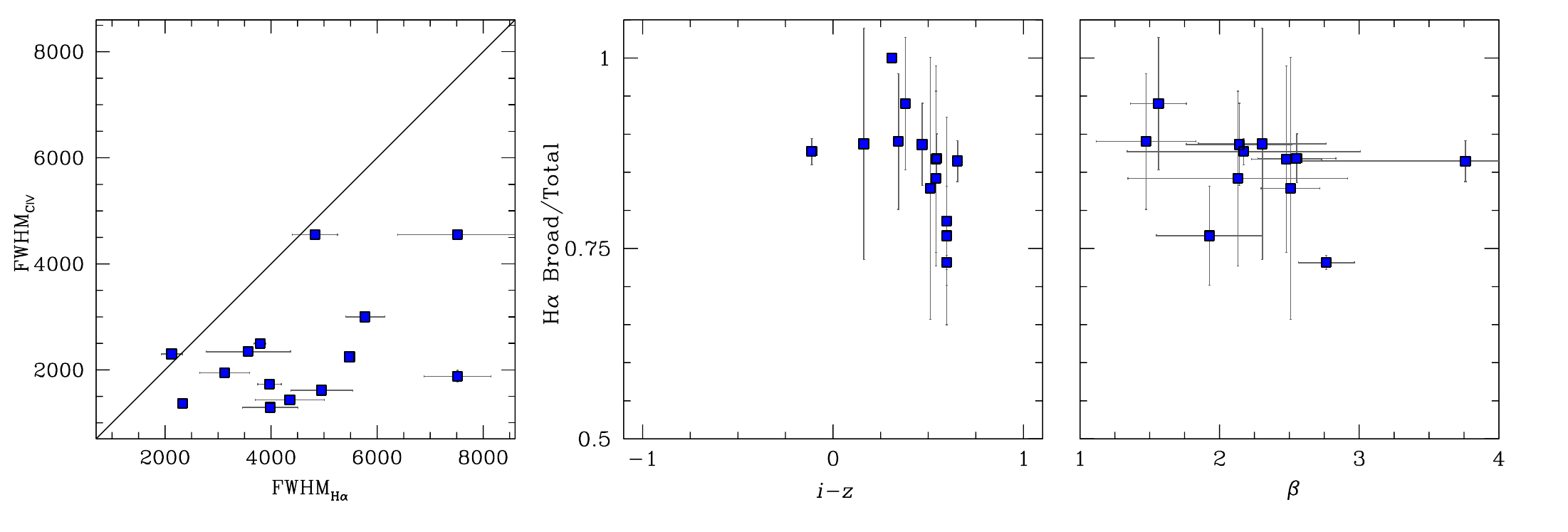}
}
\vskip -0mm
\figcaption[]{
  {\it Left}: Comparison of the broad component of the C{\tiny IV} line as
  measured from the multi-Gaussian fit with the broad component 
of H$\alpha$. Solid line indicates the unity relation. We see no clear correlation between 
the two.  The \halpha\ linewidths are systematically broader by factors of 
two or more.
{\it Middle and Right}: 
Ratio of flux in broad \halpha\ to the total flux in the line as a
function of rest-frame UV color ($i-z$; middle) and the MIR spectral
slope $\beta$ that we fit to the \emph{WISE} bands (right).  SDSS
J1406+0429, SDSS J0906+0309, and SDSS J1342$-$0045
are missing from the right-hand panel,
since they do not have sufficient \emph{WISE} detections to measure a slope.
We see no correlation between continuum color and the prominence of
the broad line, similar to the case with C{\tiny IV}.
\label{fig:fwcolorha}}
\end{figure*}

We again use the difference in $\chi^2$ of the two fits (with and
without the broad component), combined with the S/N of the line, to
decide whether the additional broad component is justified.
Absorption causes additional challenges. For example, the \ion{C}{4}
in SDSS J0812+1819 appears to have a broad base, but without the blue
side of the line our detection is not very secure.  In two cases, SDSS
J1155+0444 and SDSS J1342$-$0045, we have only upper limits on a broad
component.  Below we will examine whether the relative strengths of
the broad and narrow components depend on any interesting physical
parameters of the system.

\subsection{WISE Luminosities}
\label{sec:wise}

One of the goals of this paper is to use broad-band SED information,
including rest-frame optical and UV line luminosities and continuum
measurements, to determine the intrinsic luminosities of our
targets. Since the infrared emission from hot dust near the central
AGN should be relatively isotropic \citep[although
see][]{nenkovaetal2008,liuetal2014}, we consider the photometry from
the Wide-Field Infrared Survey Explorer (\emph{WISE}; Wright et
al. 2010), which covers 3--26~$\mu m$.  In Paper I we cross-matched
the SDSS target list with the \emph{WISE} All-Sky Data Release. We
found only a few dozen matches with $> 5 \sigma$ detections in at
least one band.  In this work, we perform forced photometry of the
\emph{WISE} images at the positions of known SDSS sources (Lang et
al., in prep). We convolve the \emph{WISE} data with the point-spread
function at the known positions of the SDSS sources.  We then fit a
model that includes a constant background level plus the point source
to a small patch ($\sim 9\times9$ pixels, or $23 \times 23$\arcsec\
for $W1-W3$ and twice that for $W4$) in the \emph{WISE} images.  The
amplitude of the point-source component of the model is a measurement
of the flux at that position in the \emph{WISE} images (regardless of
whether the source is detectable in the \emph{WISE} images).  The
formal variance of our forced photometry is equal to the per-pixel
variance of the image being photometered times the noise equivalent
area of the profile; $\textrm{var}(f) = \sigma^2 / \sum_i p_i^2$ where
$p_i$ is the normalized profile (evaluated at pixel $i$) and
$\sigma^2$ the per-pixel noise variance. We include in the fit all
images available in the AllWISE data release.

Taking this forced photometry, we can examine the colors and
luminosities of our sources in the MIR. We find that the median MIR
color of the class A sources from Paper I is $\langle W1 - W2 \rangle
= 1$ mag (Vega).  As we showed in Paper I, the colors of our targets
match those of the general BOSS quasar population at matching redshift
and luminosity.  In contrast, \emph{WISE}-based AGN selection
algorithms focus on the red end of our distribution, for instance $W1
- W2 > 0.8$ mag
\citep[e.g.,][]{sternetal2012,eisenhardtetal2012,wujetal2012}.  Our
targets are also 1-3 mag fainter in \emph{WISE} than the Stern et al.\
sample.

Next, we measure the \emph{WISE} luminosity at a fixed rest-frame
wavelength. We select $3 \micron$ as the longest wavelength that
allows inclusion of our highest redshift targets.  To find the
rest-frame $3 \micron$ luminosity, we fit all bands with a single
power law for all rest wavelengths longer than 1$\micron$, where the
quasar SED turns up due to hot dust.  We have detections in three or
more \emph{WISE} bands longward of 1$\micron$ (rest) and do not carry
out such fits in the few cases where we have a detection only in a
single band.  A simple power-law provides a fine description of the
SEDs, which are all rising towards the red.  We record both the
rest-frame 3$\micron$ flux and the slope of the power-law $\beta$ for
use below (Fig. \ref{fig:wisefit}). In the future, we will leverage
broad-band information from GALEX to \emph{WISE}, combined with
sophisticated multi-component models to the SEDs following the
technique of \citet[][]{lussoetal2013}.

\section{Results}
\label{sec:Results}

We now examine the rest-frame optical line widths and strengths in an
attempt to understand in more detail the nature of our candidate Type
II quasars.  Qualitatively, the rest-frame optical emission line
properties appear as we expect.  We find high ratios of
[\ion{O}{3}] to \hbeta\ in nearly all sources, in contrast to NLS1
galaxies that are characterized by \oiii/\hbeta$< 3$
\citep[e.g.,][]{osterbrockpogge1985}.  We detect broad components in
the \halpha\ line that are systematically broader than those we see in
\ion{C}{4}, while on average in the quasar population as a whole, \ion{C}{4}
is broader, albeit with large scatter \citep[e.g.,][]{shenliu2012}.
We only rarely detect broad \hbeta, and will use this fact to place
limits on the extinction of the broad-line region in \S \ref{sec:extinction}.

\begin{figure*}
\vbox{ 
\vskip 2mm
\hskip +20mm
\includegraphics[scale=.55,angle=0]{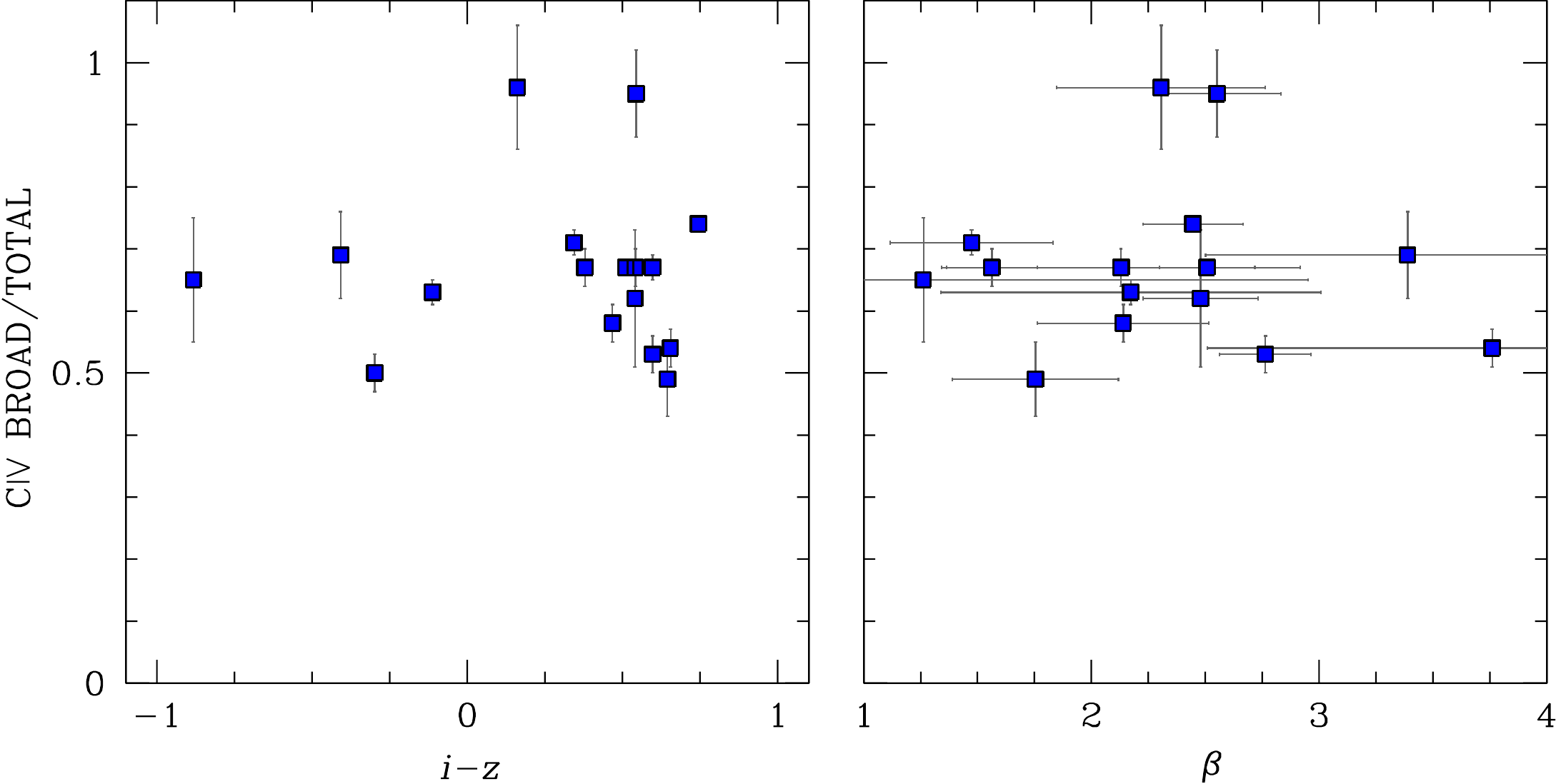}
}
\vskip -0mm
\figcaption[]{
Ratio of broad to total flux in the C {\tiny IV} line, as a function
  of rest-frame UV (left) and MIR (right) color. The narrow-line profile
  is set by our [O {\tiny III}] fits.  Broad C {\tiny IV} was detected in all cases. 
  We see no
  correlation between line width ratios and optical or mid-IR color. 
\label{fig:brcolor}}
\end{figure*}

In what follows, we will investigate the incidence of broad permitted
lines, the broad-band SEDs, and the relationship between \oiii\
rest-frame EW and luminosity as clues to the nature of our quasars.

\subsection{Line Widths and Shapes}

We start with the incidence and properties of the broad permitted
lines.  In all cases we measure line widths from the Gaussian fits.
For \oiii, we measure the FWHM from the total fit to the line. For
\halpha\ we quote the broad-line FWHM based on the Gaussian fit. In
the case of \ion{C}{4}, we calculate both a {\it total} linewidth,
measured nonparametrically on the combined narrow+broad line fit, for
comparison with \oiii, and a {\it broad} linewidth using just the
single broad Gaussian, for comparison with \halpha.  Recall that
  these targets were originally selected to have \ion{C}{4} line
  widths $< 2000$~\kms\ based on a single Gaussian fit with the width
  tied to other emission lines (excluding Ly~$\alpha$) performed by the
  SDSS pipeline \citep{boltonetal2012}.  The line widths that we
  present here make the physically motivated assumption that the
  \oiii\ linewidth traces the narrow-line region, while any
  additional component we find in the \ion{C}{4} lines arises from the
  broad-line region.  Thus we are able to uncover weak broad
  emission not captured by the pipeline fit.

Narrow-line widths, as measured from the \oiii\ lines, range from 200
to 1000 \kms, consistent with lower-redshift narrow-line regions in
luminous targets \citep[Fig. \ref{fig:fwcolor},
e.g.,][]{haoetal2005,reyesetal2008,liuetal2013b}.  We detect
significant broad \halpha\ in all of the targets, ranging in width
from 1000 to 7500~\kms.  Likewise, we detect an additional component
beyond the \oiii\ model in every \ion{C}{4} line.  The total
\ion{C}{4} linewidths span only $1000$ to $2000$~\kms (Figures
\ref{fig:nirfits} \& \ref{fig:nirfitsAPO}).  The broad \ion{C}{4}
linewidths range from 1300 to 4500~\kms, but nearly all have FWHM
$\lesssim 3000$~\kms, and they are systematically lower than the broad
component seen in \halpha\ (Figure \ref{fig:fwcolorha}, left).

We seek correlations between the broad-line widths and broad-line
fractions with continuum color.  First, in Figure \ref{fig:fwcolor},
we compare the profile widths of \oiii\ (the bona-fide narrow-line
region) and total linewidth of \ion{C}{4}.  The two are strongly
correlated, but the \ion{C}{4} line is $\sim 2.5$ times as broad as
\oiii\ on average. We see no clear correlation between FWHM$_{\rm
  [O{\tiny III}]}$/FWHM$_{\rm CIV}$ and optical or mid-infrared
color. We also search for a correlation in line shifts between \oiii\
and \ion{C}{4}, but found no correlation (Spearman rank $\rho=0.15$,
probability of no correlation $P=0.5$).

To emphasize the narrowness of the broad component of \ion{C}{4}, we
compare it with the broad \halpha\ (Figure \ref{fig:fwcolorha},
left). With a couple of exceptions, the \halpha\ lines are roughly
twice as broad as \ion{C}{4}.  The ratio of \ion{C}{4} to \halpha\ in
typical blue quasars is known to span a large range
\citep[e.g.,][]{greeneetal2010,assefetal2011,shenliu2012,hoetal2012},
but on average the \ion{C}{4} line is broader than \halpha\ in
unobscured quasars.  We also examine the ratio of broad to total flux
in \halpha\ (Figure \ref{fig:fwcolorha}) and \ion{C}{4} (Figure
\ref{fig:brcolor}).  The broad component dominates the total line flux
in both lines in most cases.  There is no strong correlation between
the broad-line fraction and mid-infrared color, perhaps because the
latter spans such a narrow range.

Our targets were selected to have narrow \ion{C}{4}, but with
physically motivated decompositions, we can state that there is a
broad component distinct from \oiii\ and that it is relatively narrow.
While most of our targets have broad linewidths $\sim 2000$~\kms\
(Figure \ref{fig:fwcolorha}), in the \citet{shenetal2011} compilation
of SDSS quasars, only 1\% of $2 < z < 3$ quasars have \ion{C}{4} lines
narrower than $2000$~\kms. The fraction rises to 10\% when we match in
luminosity ($-23.5 < M_i < -25.5$ mag, with no extinction
  correction applied to either sample). In contrast, comparing the
distribution of \halpha\ linewidths between our sample and the objects
in the \citet{shenetal2011} catalog with $-23.5 < M_i < -25.5$ mag, we
find comparable median FWHM in \halpha\ of $4000$~\kms\ and 3300~\kms,
respectively.  A Kolmogorov-Smirnov test suggests the two
distributions are marginally consistent ($P=0.06$), with our objects
having broader lines on average. Thus, extinction is a natural
culprit to explain the weakened broad emission lines at UV
wavelengths, assuming a standard wavelength dependence for 
the reddening law. In the next section we will quantify the amount of
extinction needed to explain our observations. We note that some of the absorption
may arise from intervening \citep{yorketal2006} or associated 
absorbers \citep{vandenberketal2008}.

\begin{figure*}
\vbox{ 
\vskip +5mm
\hskip +10mm
\includegraphics[scale=.65,angle=0]{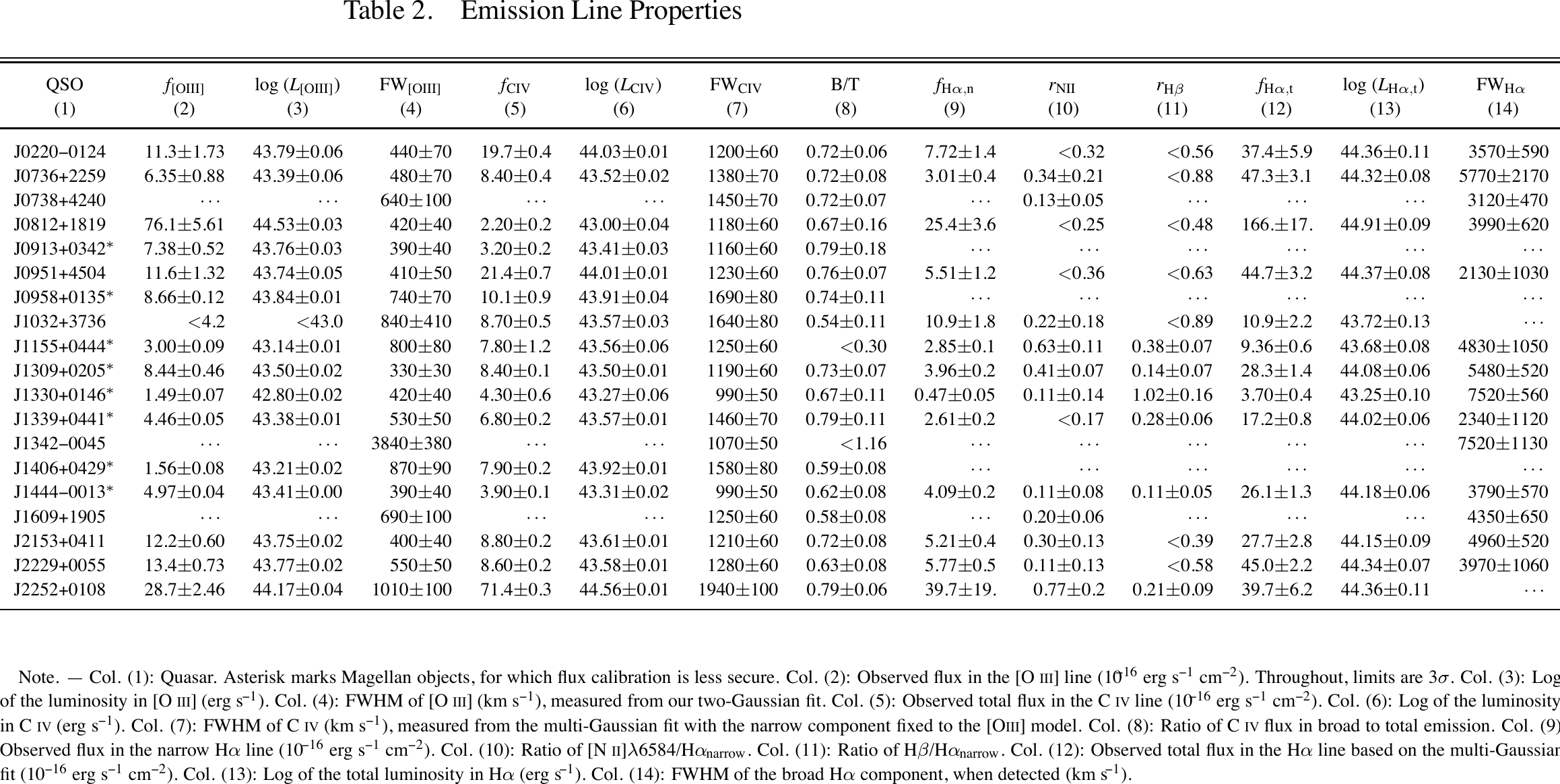}
}
\vskip -0mm
\label{tab:line}
\end{figure*}
\vskip 5mm

To summarize this section, we detect permitted line emission that is
broader than the narrow-line model alone.  The broad \ion{C}{4} lines
are relatively narrow (with FWHM ranging from $1000$ to $4500$~\kms,
but most having FWHM$\approx 2000$~\kms). These relatively narrow
  linewidths are not surprising, given the initial sample selection.
In contrast, the \halpha\ lines span a more typical range of widths
from $1000$ to $7500$~\kms.  Let us now examine whether we can explain
the difference between these two lines using extinction.

\subsection{Equivalent Widths and Extinction Estimates}
\label{sec:extinction}

Another valuable clue to the nature of our sources comes from
comparing the \oiii\ line rest-frame equivalent widths (EWs) and
luminosities.  We assume that the \oiii\ is extended relative to the
broad-line region. Moving from unobscured to obscured sources, we
would expect $L_{\rm [O III]}$ to stay constant at a given bolometric 
luminosity while the continuum is
obscured and grows fainter. As a result, obscured sources should have
much higher EWs at a given line luminosity (see Figure
\ref{fig:ewo3}). The observed range of the \oiii$~\lambda
5007$\AA\ line EW is bracketed by obscured
quasars at $0.5 < z < 0.8$ from \citet{zakamskaetal2003} and
unobscured quasars at comparable redshift from
\citet{shenetal2011}. Our targets lie between the
obscured and unobscured sources. 

We can derive an estimate for the extinction using the ratio of the
observed rest-frame EW to that expected for an unobscured quasar with
the same $L_{\rm [O III]}$.  Since we expect the \oiii\ flux to emerge
from larger scales with lower extinction, the ratio of the observed
and expected EW simply diagnoses the extinction level of the continuum
at 5007~\AA.  Assuming SMC-like dust \citep{pei1992}, which appears
appropriate for quasars \citep{hopkinsetal2004}, we calculate the
extinction in the $B$-band, $A_B$, for the seven quasars with
EW$>$EW$_{\rm unobscured}$ and an \halpha\ measurement.  Note
  that this estimate is independent of the absolute flux calibration.
Based on this calculation, we find values of $A_B$ ranging from $0.07$
to $2.8$ mags with a median of $A_B = 1.2$ mag (Table 3; $0.05 <
  A_V < 2.2$ mag; median $A_V = 0.9$ mag).  For reference, a value of
$A_B = 1.2$ mag corresponds to a transmitted UV flux at 1500\AA\ that
is 14\% of the intrinsic flux, assuming the SMC reddening curve of
\citet{pei1992}. The two largest extinction values belong to SDSS
J1330+0146 and SDSS J0958+0135 
\citep[the latter is in COSMOS;][Paper I]{scovilleetal2007}.

Given that in general we detect broad \halpha\ but not broad \hbeta, we can
derive a second estimate of the reddening $A_B$ if we make the assumption
that all objects have the same intrinsic \hbeta/\halpha\ ratio. We
derive $3\sigma$ limits to the broad \hbeta\ flux by integrating the
\hbeta\ fit residuals over twice the FWHM of the \halpha\ line. We
then estimate a lower limit to the extinction, assuming an intrinsic
Case B$^{\prime}$ \halpha/\hbeta\ ratio of 3:1.  This ratio is reasonable, 
since the typical observed \halpha/\hbeta\ ratio in lower redshift SDSS quasars is 
$4:1$ \citep{greeneho2005cal}.

Going a step further, we predict the broad \ion{C}{4} flux based on
the observed broad \halpha\ flux and the calculated extinction.  We
assume that intrinsically the \ion{C}{4} line is 35\% brighter than the \halpha\
line, based on the compilation of \citet{shenliu2012} of 60 SDSS
quasars with $1.5 < z < 2.2$.  We will use this single value although
there is nearly an order of magnitude scatter in the 
ratio\footnote{Our observed C{\small IV}/H$\alpha$ ratios 
range from 0.04 to 2. We find no significant correlation between 
the observed line ratios and the reddening, but the total number of 
objects is small.}.  
The expected broad \ion{C}{4} flux is nearly always higher than
what is observed by a factor of two to ten
(with measurement errors causing a 
factor of $\sim 2.5$ uncertainties on these ratios).  The one
exception is SDSS J1330+0146, where the expected flux is only 6\% of
what is observed, but the broad emission lines are both 
quite weak in this case. There are two possible interpretations.  Given that
the broad component of \ion{C}{4} is much narrower than that of
\halpha, one possibility is that we are not detecting the full broad
\ion{C}{4} line, and that the actual extinction is greater than what
we estimate from the Balmer decrement.  Alternatively, the extinction
curve may be flatter in the UV than the SMC curve
\citep[e.g.,][]{gaskelletal2004}, in which case we would be overpredicting
the intrinsic \ion{C}{4} luminosity by factors of several. A third possibility is 
that the C{\small IV}/H$\alpha$ ratios are anomalous in these objects.

\begin{figure*}
\vbox{ 
\vskip +15mm
\hskip +15mm
\includegraphics[scale=0.6,angle=0]{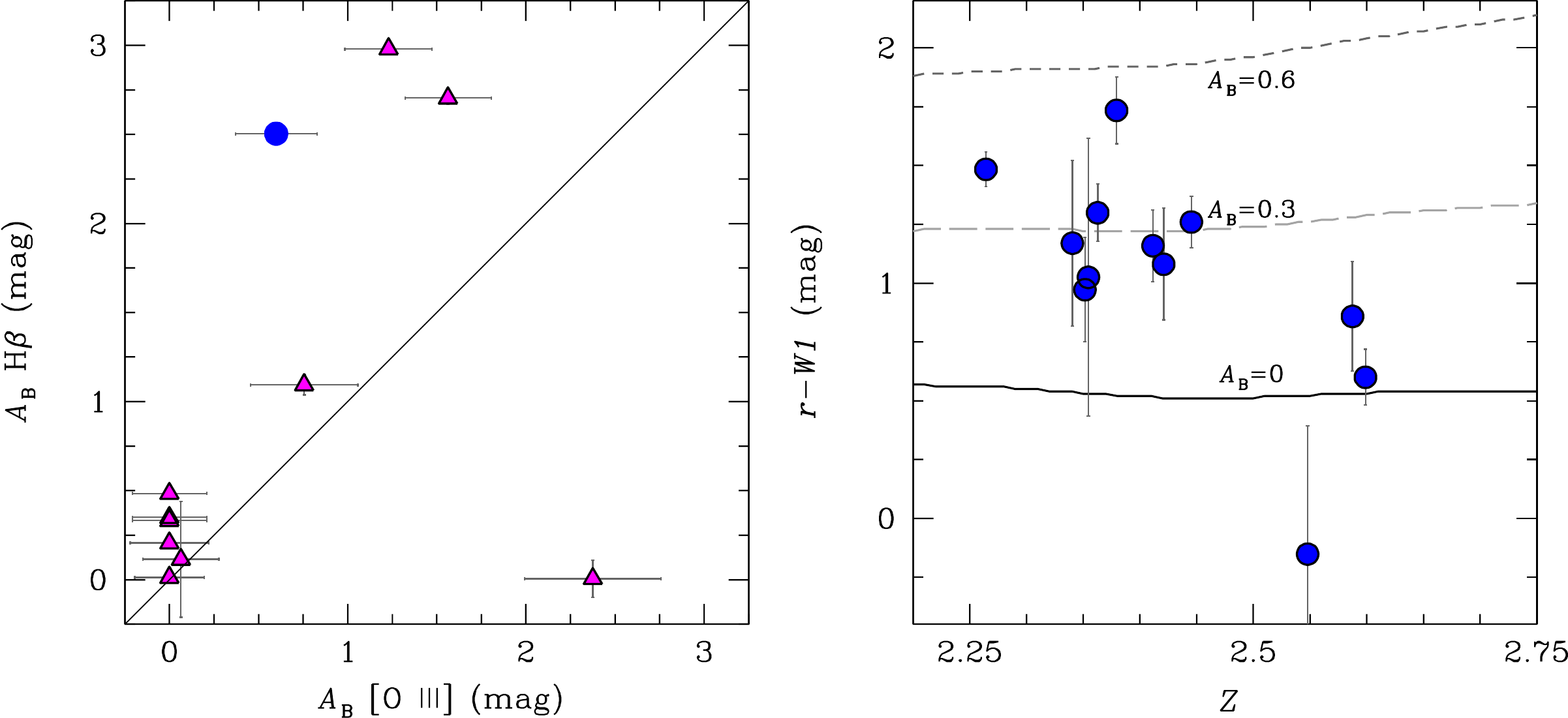}
}
\vskip -0mm
\figcaption[]{
{\it Left}: We compare two estimates for $A_{B}$ (mag).  The first is derived by
assuming that any positive deviations from the average [O {\tiny III}]
EW are caused by extinction of the continuum. The second takes limits
on the broad \hbeta\ flux and assumes an intrinsic \halpha/\hbeta\
ratio of 3.01:1.  The latter estimate is a lower limit on $A_B$, to
the extent that the assumption of Case B$^{\prime}$ ratio is correct.
The blue symbol is SDSS J1309+0205, for which we detect a broad
\hbeta\ component. The solid line indicates the 1:1 relation.
{\it Right}: The observed $r-W1$ color as a function of redshift, plotted against 
the color of the \citet{richardsetal2006} SED with absorption ranging from 
$A_{B} = 0$ to $A_{B} = 0.6$ mag, showing that only moderate extinction 
is required to explain the observed UV to NIR colors.
\label{fig:civhbab}}
\end{figure*}

We compare the EW estimate for $A_B$ with the limits derived from the
flux ratio of \hbeta/\halpha\ in Figure \ref{fig:civhbab} (left).  A
value of $A_B = 0$ on the x-axis means that the \oiii\ EW is
consistent with the unobscured quasars. The two estimates are
correlated, which is encouraging, since both are estimates of the
reddening to the continuum/broad-line region.  In general, the
\hbeta-based absorption estimates are larger, perhaps because the
underlying \hbeta\ to \halpha\ ratios span a wide range, rather than
having the Case B$^{\prime}$ recombination value
\citep[e.g.,][]{macalpine1985,rixetal1990,koristagoad2004,greeneho2005cal}.
The other possibility is that the \oiii-derived values are too low,
because we have ignored extinction of the \oiii\ line and/or scattered
light boosting the continuum emission in the Type II quasars.  As a
sanity check, we can also calculate the minimum extinction needed to
bring the $r-W1$ color in line with the \citet{richardsetal2006} blue
quasar SED.  We find this minimum to be $A_B < 0.5$ mag ($A_V < 0.4$
mag) in all cases, consistent with the other estimates but less
restrictive (Figure \ref{fig:civhbab}, right).  We do not calculate
the extinction using the narrow \hbeta\ and \halpha\ lines because of
the low significance of the \hbeta\ detections.

In summary, based on the EW of \oiii\ and limits on the Balmer
decrement in the broad emission lines, we find evidence for modest 
absorption ($A_B \approx 0-3$ mag or $A_V \approx 0-2.2$ mag).

\subsection{Luminosity Indicators}

We now investigate the SEDs and bolometric luminosities of these
quasars.  In a future paper we will present full SEDs, but for the
moment we simply compare rest-frame UV and optical line luminosities
with UV and mid-infrared continuum luminosities.  In the case of
unobscured sources, the rest-frame UV light traces the big blue bump
directly, while the near to mid-infrared emission (longward of $\sim 1
~\micron$) traces hot dust sitting close to the accreting black
hole. 

\vbox{ 
\vskip +1mm
\hskip -2mm
\includegraphics[scale=0.4,angle=0]{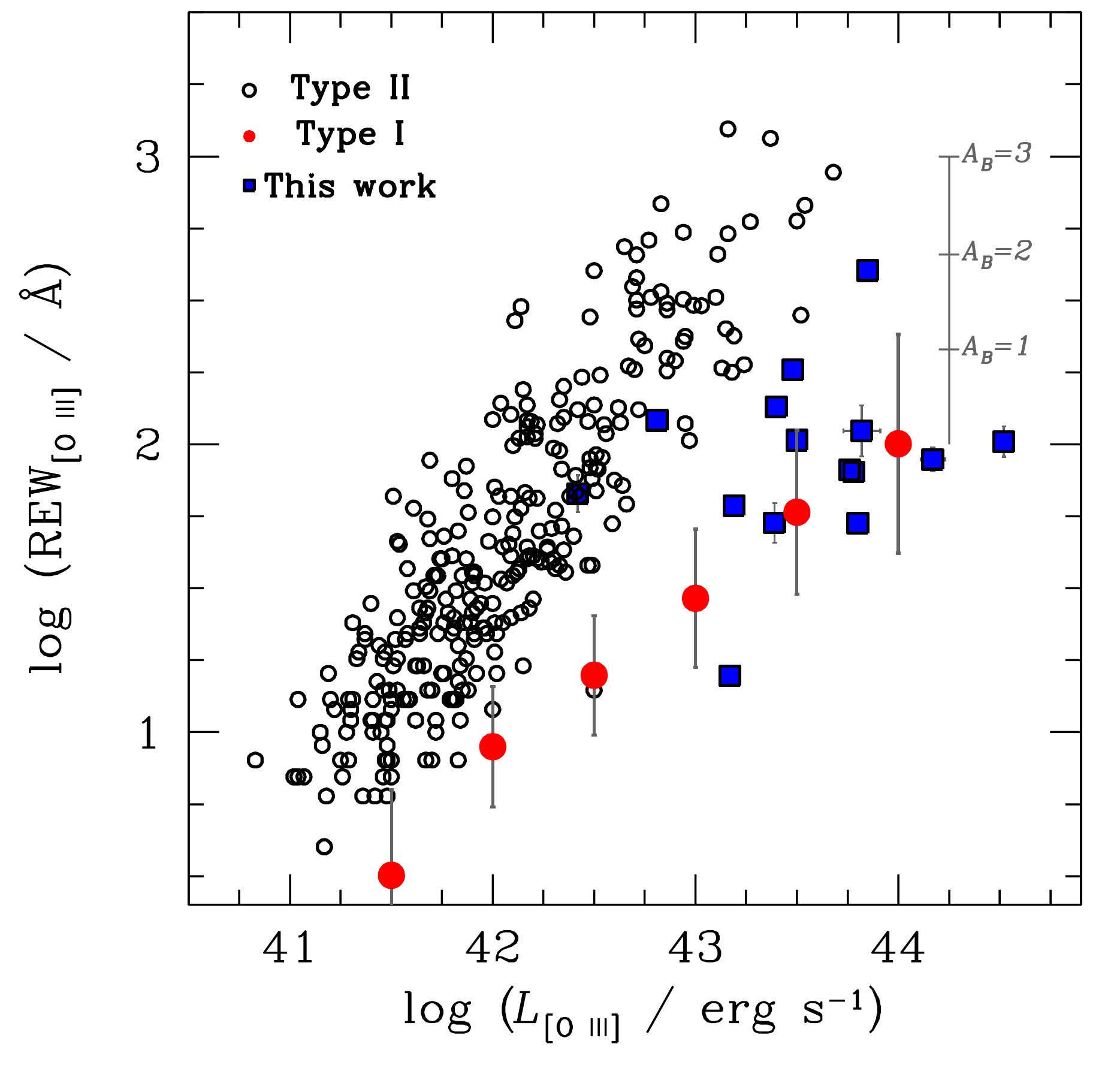}
}
\vskip -0mm
\figcaption[]{
The rest-frame 
equivalent width (EW) of [O {\tiny III}] compared with the luminosity in the 
same line. The blue squares are the high-redshift obscured quasar candidates 
presented here.  The open black circles are Type II quasars at lower 
redshift from \citet{zakamskaetal2003}, while the red circles are median 
values from the \citet{shenetal2011} catalog of Type I objects with $0.5 < z < 0.8$ 
to roughly match the Zakamska sample. Our objects are intermediate in 
properties between the obscured and unobscured sources. The 
extinction needed to make the [O {\tiny III}] EWs commensurate with the Type I 
objects' $A_{B}$ is between 0 and 3 mag, as shown schematically with the grey bar. 
Two points are missing here, SDSS J0951+4504, where we only marginally 
detect the continuum, and SDSS J1032+3736, where we do not detect 
the [O {\tiny III}] line.
\label{fig:ewo3}}
\vskip 5mm

\begin{figure*}
\vbox{ 
\hskip +25mm
\includegraphics[scale=0.95,angle=0]{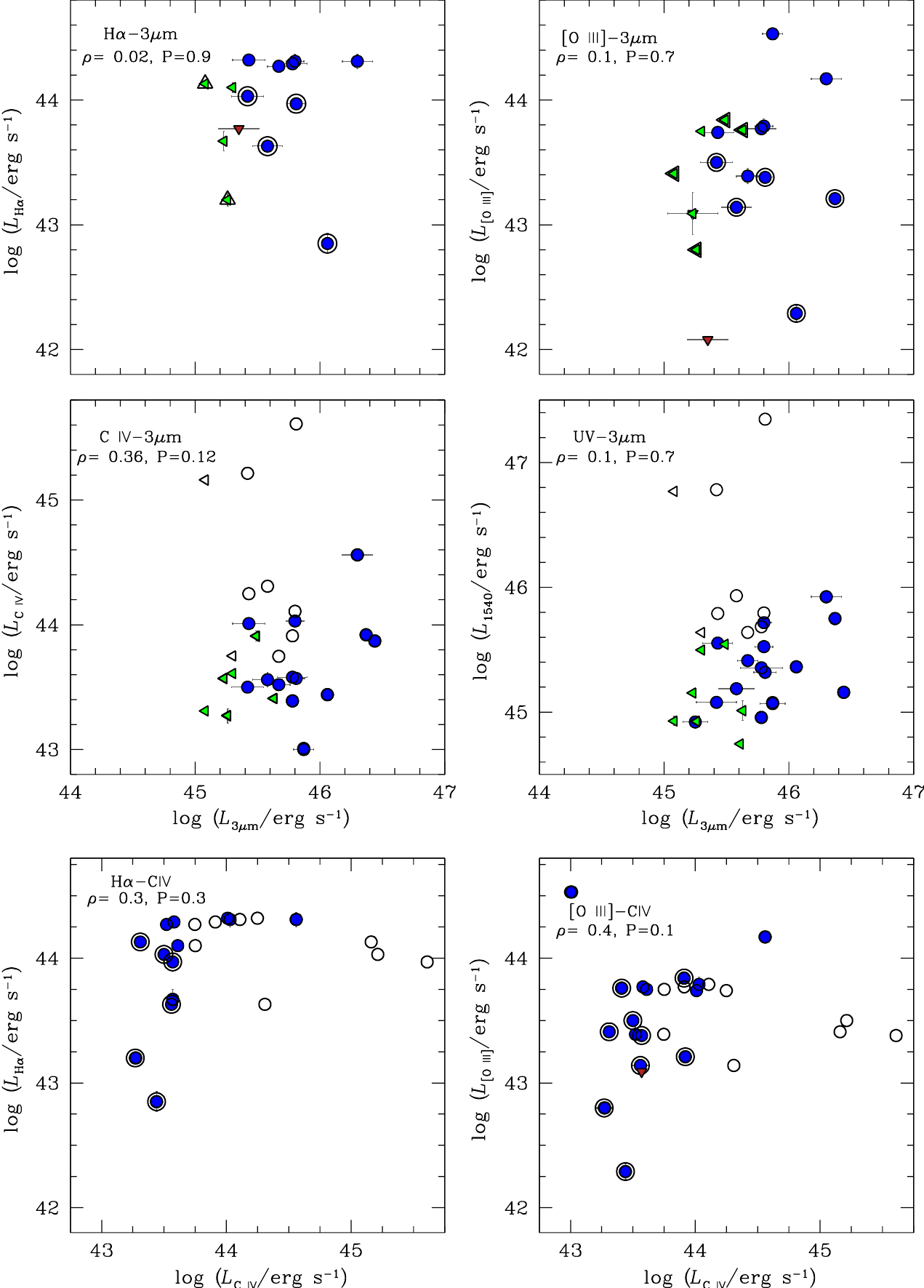}
}
\vskip -0mm
\figcaption[]{
  Comparison of various luminosity indicators, all in units of
  erg~s$^{-1}$.  The bottom two panels compare the line luminosity of
  C {\tiny IV} with total \halpha\ (left) and [O {\tiny III}]
  (right). Magellan targets are indicated with double symbols; we see
  that while we preferentially targeted sources with lower rest-frame
  optical-UV luminosities with Magellan, these targets alone are not
  responsible for the observed scatter.  In the top four panels, we
  compare with the \emph{WISE} luminosity at a rest-frame wavelength
  of 3\micron.  Errors are plotted on all points, but are often
  comparable to the symbol size. The $y$-axis $3\sigma$ upper limits
  are plotted as downward red arrows, while the $x$-axis $3\sigma$
  upper limits are plotted as green rightward pointing arrows. The
  dereddened values are shown as open symbols and are based on the
  Balmer decrement limits from the non-detection of broad \hbeta\
  lines. There are no strong correlations between any of these
  indicators, although in general they cover comparable ranges in
  luminosity. The one exception is the narrow range of the UV
  continuum luminosity, perhaps just reflecting our UV selection.
\label{fig:wiselum}}
\end{figure*}

\noindent
The latter, therefore, is expected to be relatively isotropic
\citep[although see also][]{nenkovaetal2008,liuetal2014}. Line
luminosities, in so far as they are photoionized by the central
source, also correlate with the continuum luminosity
\citep[e.g.,][]{yee1980,zakamskaetal2003,
  heckmanetal2004,greeneho2005cal,liuetal2009}.  The narrow line
luminosities should also be relatively isotropic \citep[although see
also][]{netzeretal2005}.  We compare various
intrinsic luminosity indicators with each other in Figure
\ref{fig:wiselum}.  In the case of \halpha\ and \ion{C}{4}, we plot
total (broad+narrow) luminosities.  We do this for consistency with
other work \citep{greeneho2005cal}, but since the narrow emission is
such a small fraction of the total flux, removing it would make little
difference to the final outcome.

\vskip 5mm

\vskip 5mm

\vbox{ 
\vskip -5mm
\hskip +1mm
\includegraphics[scale=0.65,angle=0]{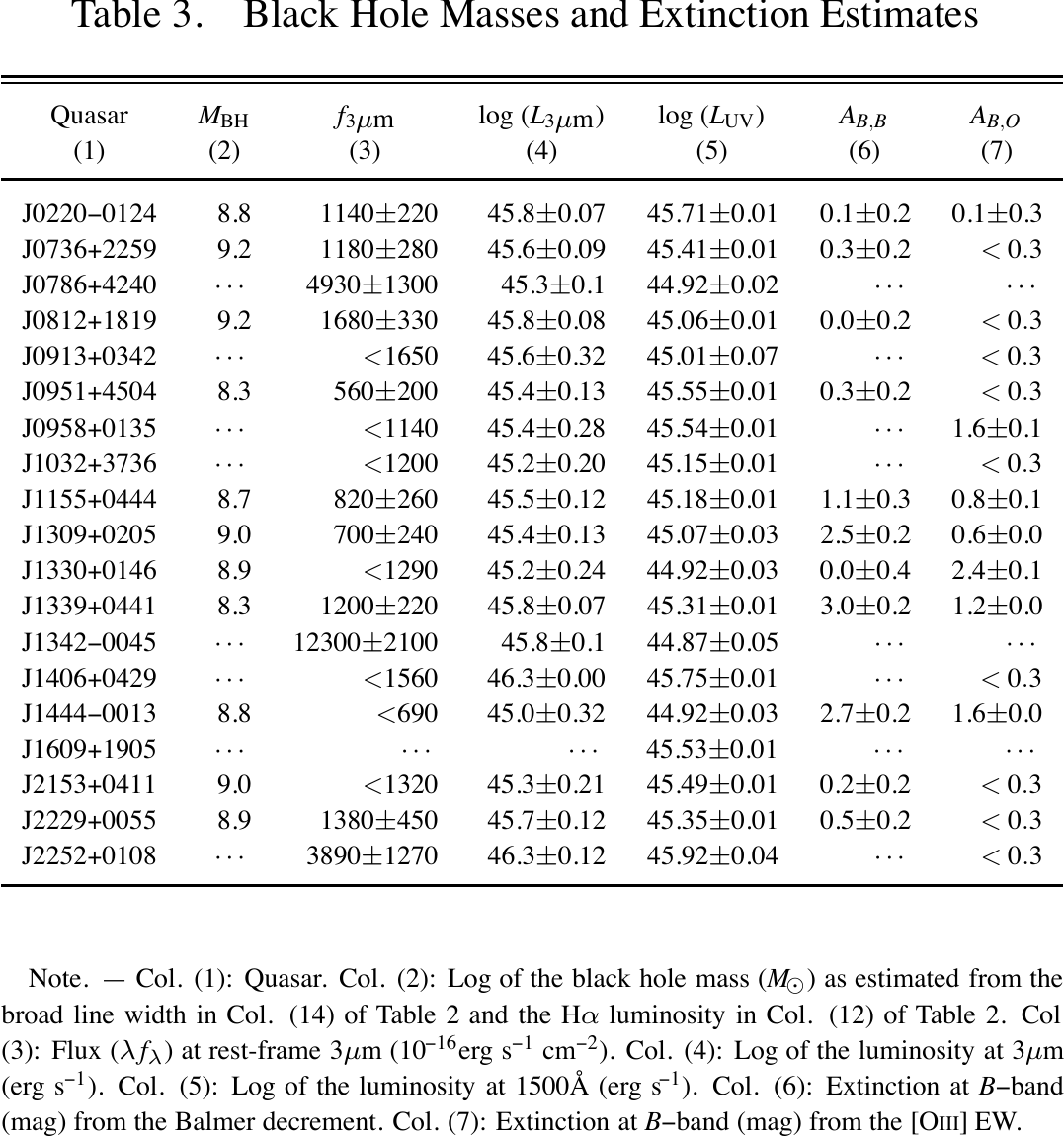}
}
\vskip -0mm
\label{tab:line}
\vskip 5mm

If there is substantial extinction of the broad-line region
specifically, then we expect the weakest correlation between the
strength of the \ion{C}{4} line and the UV continuum emission, while
correlations with progressively redder diagnostics should be stronger.
The tightest correlation is expected to be between the \emph{WISE} and
\oiii\ luminosities, as the most isotropic.

In fact, we do not find significant correlations between any of our
luminosity probes (Figure \ref{fig:wiselum}), based on Spearman rank
coefficients.  We have also investigated $L_{\rm H\alpha}$ and $L_{\rm
  [OIII]}$ against the UV luminosity, and find no significant
correlation. We also show the ``dereddened'' UV and \ion{C}{4} fluxes,
adopting the Balmer decrement estimates of $A_{\lambda}$; even
with an extinction correction, we detect no correlation.  In general
the luminosity indicators all span a similar range of
$\sim 2$ dex.  The one exception is the observed UV luminosity, which
spans a narrow range. Considering that we selected the targets to
be at the flux limit of the BOSS survey and to have high-EW emission
lines, the narrow observed range in UV luminosity may be no more than a selection
effect.

There are two possible explanations for the lack of luminosity
correlations.  One is the limited range in observed luminosity of the
sample.  It may be that intrinsic SED differences, combined with
extinction of the narrow-line region, wash out correlations over the
narrow luminosity range we probe here. Bona fide Type II quasars at $z
\sim 0.5$ are clearly redder in the MIR than are unobscured quasars,
possibly adding additional scatter to the \emph{WISE} luminosities
\citep[e.g.,][]{liuetal2013b}.  The other major caveat here remains
outstanding uncertainties in our flux calibration.  We have tried
splitting the data into APO and Magellan subsamples to see if a strong
correlation holds with one or the other, but the total number of
objects is too small to seriously address this question.

\subsection{Black Hole Masses}

With rest-frame optical spectra, and detected broad emission lines, we
can use scaling relations to estimate black hole (BH) masses for these
luminous quasars
\citep[e.g.,][]{vestergaard2002,shenetal2008,kellyetal2009}.  We
assume that only gravity influences the motions of the broad-line
region gas, so that the velocity dispersion of the gas relates
directly to \mbh; $\mbh \propto R \upsilon^2$.  We also need a size
scale for the emitting gas, which we estimate from the AGN luminosity
using the ``radius-luminosity'' relation, calibrated using
radius measurements from reverberation mapping
\citep[e.g.,][]{kaspietal2000,bentzetal2009,bentzetal2013}.  Because
we do not detect the optical AGN continuum directly in our
observations, we rely on the known correlation between continuum and
line luminosity \citep[e.g.,][]{yee1980} to calculate the BH mass from
observations of the \halpha\ line alone (uncorrected for extinction).  
The relationship between
continuum luminosity and line luminosity, as well as the relation
between \hbeta\ and \halpha\ FWHM, are taken from
\citet{greeneho2005cal}, while the radius-luminosity relation is taken
from \citet{bentzetal2013}.

BH masses derived in this manner clearly carry a large number of
systematic uncertainties that are difficult to quantify
\citep[e.g.,][]{krolik2001,greeneho2006}. Including these
systematics, the BH mass uncertainties are estimated to be factors of
a few \citep[e.g.,][]{vestergaardpeterson2006,shen2013}.  Eventually
we hope two-dimensional reverberation mapping of nearby Seyfert
galaxies will mitigate these problems
\citep[e.g.,][]{bentzetal2008,denneyetal2010,barthetal2011,
  pancoastetal2012,kashietal2013}, but in the meantime we must be
careful not to overinterpret the BH mass estimates.

Taking the measurements at face value, the derived BH masses are
typically $\sim 10^9$~\msun.  Technically, these are lower-limits,
since they rely on the \halpha\ luminosity, which almost certainly
suffers extinction.  On the other hand, the inferred \mbh\ only depends
on the square root of the \halpha\ luminosity; a factor of three error
in broad \halpha\ luminosity corresponds to only a $0.24$ dex change
in \mbh.  The line width may be underestimated due to extinction as
well, but given that the distribution of \halpha\ widths spans a
similar range to other quasar samples
\citep[e.g.,][]{greeneho2007,shenliu2012,banerjietal2012,
  matsuokaetal2013,banerjietal2013}, it is hard to estimate how large
that bias may be. For comparison, we show both the median source
  from the SDSS sample in a matching redshift and magnitude range
  \citep{shenetal2011} as well as the median object from a
  NIR-selected sample presented in \citet{banerjietal2012}.

We can ask whether these masses are reasonable given the total
luminosities of the objects.  In addition to factors of a few
uncertainties in \mbh, there are also a number of uncertainties
involved in calculating the bolometric luminosities.  The composite
SED of unobscured objects in \citet{richardsetal2006} shows that the
bolometric luminosity is $7-10$ times $L_{3 \micron}$. Using this
scaling, the majority of the sample has sensible Eddington ratio
estimates of $\sim 10\%$, with all targets consistent with being below
the Eddington limit.  Considering the lack of correlation between
\emph{WISE} and \oiii\ luminosities discussed above, these Eddington
ratios are no better than order-of-magnitude estimates.  The largest
uncertainty is in our choice of the broad emission line used to 
measure the gas velocity dispersion.  If we used the
\ion{C}{4} linewidth instead of the \halpha\ linewidth, our \mbh\
values would be $\sim 4$ times smaller on average.  However, 
there may be a component of the \ion{C}{4} that is
completely obscured. Thus the \halpha\ linewidth may be more
representative of the broad-line region kinematics.

\vbox{ 
\vskip +1mm
\hskip -2mm
\includegraphics[scale=0.45,angle=0]{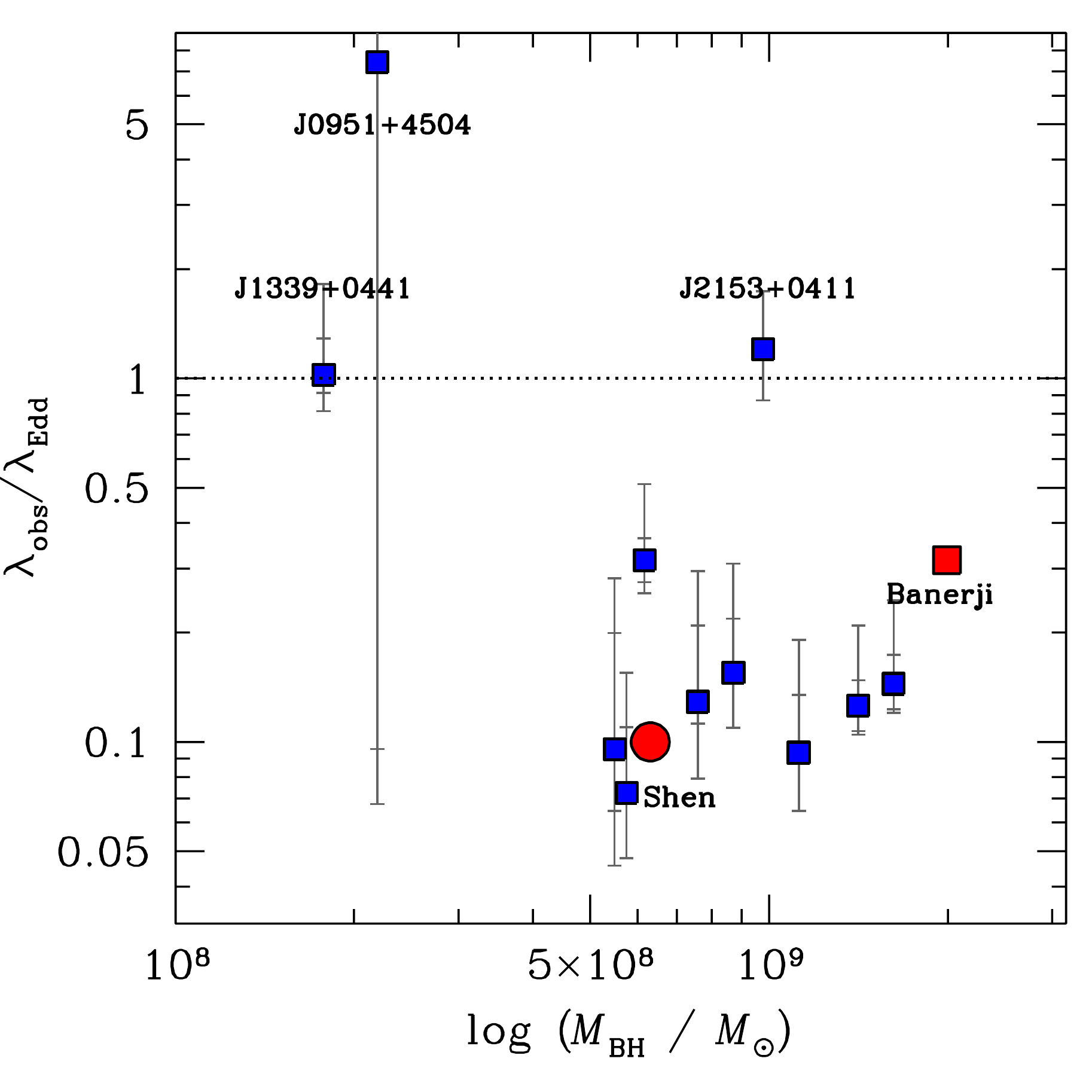}
}
\vskip -0mm
\figcaption[]{
BH masses compared with the ratio of bolometric to Eddington
  luminosity, $\lambda$. We estimate BH masses based on the \halpha\
  FWHM and the \halpha\ luminosity \citep{greeneho2005cal}. The
  bolometric luminosities are derived from $L_{\rm 3 \micron}$, and
  the uncertainties reflect both random errors (in \halpha\ 
  luminosity and FWHM) and
  systematic errors in bolometric corrections, as indicated by the double 
  error bar.   Systems with very large error bars have very uncertain 
  broad \halpha\ FWHM. We have {\it not}
  incorporated systematic uncertainties in the BH masses here.  Without that
  uncertainty, we can already see that the majority of the targets do not
  violate the Eddington limit ($\lambda =1$).  For comparison, we show the median 
\mbh\ and $\lambda$ for the optically selected sample of \citet{shenetal2011} 
in a matching redshift and magnitude range (big red circle), as well as the median 
\mbh\ and $\lambda$ for the NIR-selected sample of \citet{banerjietal2012}.
\label{fig:bhledd}}
\vskip 5mm

\subsection{Multi-peaked Objects}
\label{sec:Multi}

\begin{figure*}
\vbox{ 
\vskip -7mm
\hskip +40mm
\includegraphics[scale=0.85,angle=0]{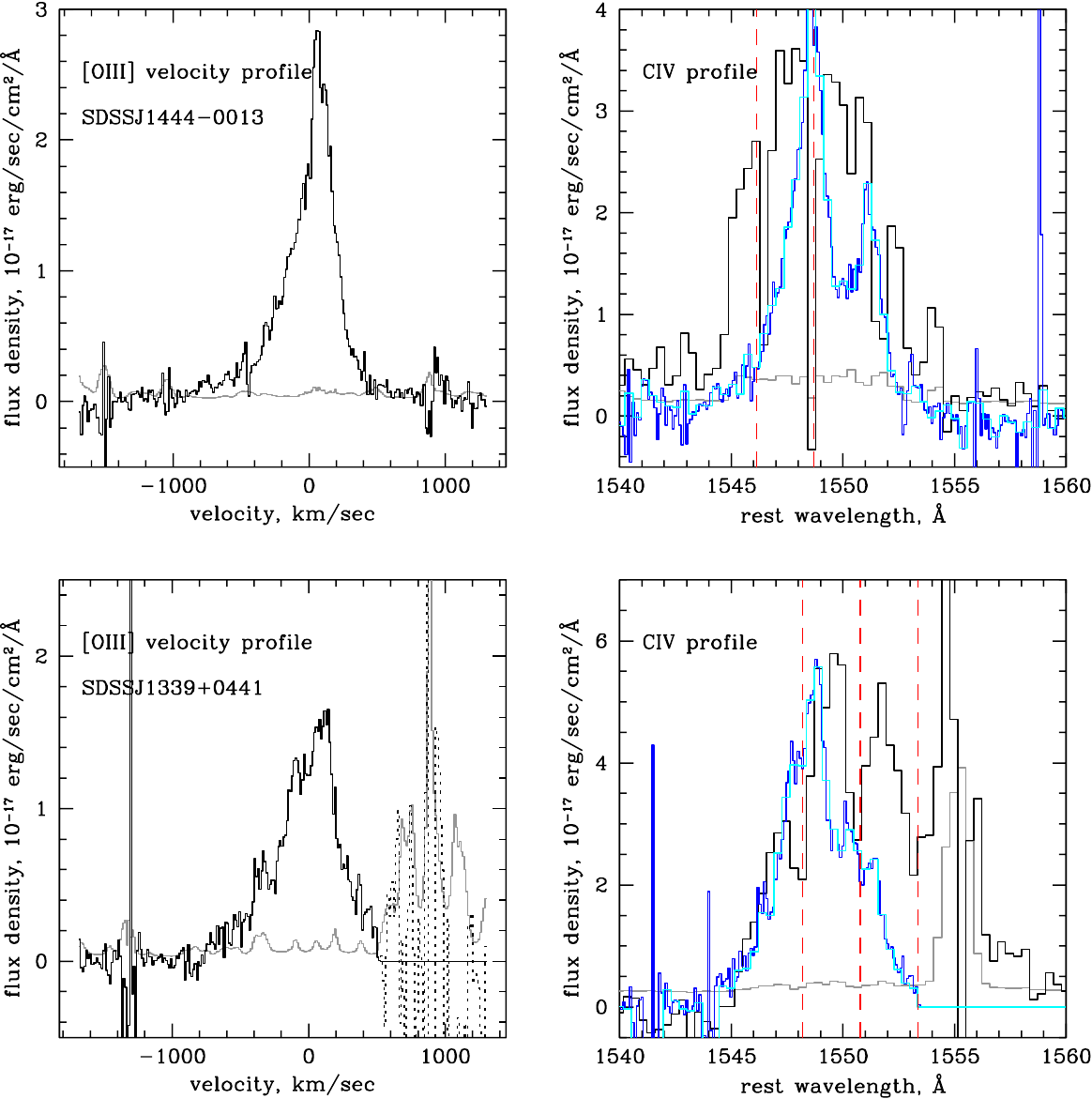}
}
\vskip -0mm
\figcaption[]{
The  [O {\tiny III}]$~\lambda 5007$ and C {\tiny IV} profiles for two 
  objects with multiple C {\tiny IV}  velocity peaks.
  We attempt to model the C {\tiny IV} doublet using the velocity
  profile of the [O {\tiny III}] line. On the left we show the [O
  {\tiny III}] line, while on the right we superpose the [O {\tiny
    III}] model on the C {\tiny IV} spectrum, with the two components
  fixed to the wavelengths of the doublet and a line ratio of 1:2 
  \citep{hamannetal1995}, although we tried 0.9:1 as above 
  and it did not change the results.  We present the Magellan spectrum (dark blue), 
  the Magellan spectrum smoothed to BOSS resolution (light blue), 
  and the error array (grey). Differing spectral resolution 
  alone cannot explain the observed discrepancy in line shape. In
  the lower left panel, the high noise immediately to the red of the
  [O {\tiny III}] line in SDSS~J1339+0441 shown with the dotted line is due to
  atmospheric emission residuals. We mask this part of the [O {\tiny III}]
  velocity profile in our model of the
  C {\tiny IV} profile.  Although velocity structure is clearly visible in the 
  [O {\tiny III}] line for SDSS
  J1339+0441, we cannot fully explain the C {\tiny IV} profile in
  either source. For example, the C {\tiny IV} emission is both
  broader and more complex than the [O {\tiny III}] emission in both
  cases.  Furthermore, there is an apparent additional 
  absorption system in both cases (red dotted lines), 
but these putative absorption systems 
  would be very narrow, and do not have the proper line ratios between the two 
  C {\tiny IV} components.
\label{fig:onedclump}}
\end{figure*}

Paper I highlighted an intriguing sub-sample of our targets that
contained multiple velocity peaks in their \ion{C}{4} spectra.  We
postulated that, since \ion{C}{4} is a resonance line, absorption was
the most likely explanation for the subcomponents, rather than
physically distinct clumps of gas with differing kinematics. We specifically 
obtained NIR spectra for two of these sources (SDSS J1339+0441 and 
SDSS J1444$-$0013, the latter subsequently fell out of the main sample 
in Paper I due to revised \ion{C}{4} line width measurements). Here we 
compare the \ion{C}{4} and \oiii\ line profiles (Figure \ref{fig:onedclump}). 

In Figure \ref{fig:twodclump} we show the two-dimensional,
flat-fielded and sky-subtracted FIRE spectrum of SDSS J1339+0441,
which clearly has a multi-peaked structure in both the 4959 and 5007
lines.  We also see strong sky residuals directly to the red of
\oiii$~\lambda 5007$.  In SDSS J1339+0441 the \oiii\ line has a
complicated velocity structure with multiple kinematic components,
although less distinct than that in \ion{C}{4}. This profile could be
due to outflow kinematics
\citep{greeneho2005gas,greeneetal2011,liuetal2013a,liuetal2013b} or to
narrow line regions around distinct black holes
\citep[e.g.,][]{djorgovskietal2007,barthetal2008,comerfordetal2009,
  liuetal2010a,liuetal2011}, as found in low-redshift objects. Another
intriguing possibility specific to high-redshift objects is that the
quasar host galaxy is being formed from multiple components which are
illuminated by the main quasar
\citep[e.g.,][]{elmegreenetal2008,forsterschreiberetal2011}.

\vskip 5mm

We now directly compare the velocity structure in the \ion{C}{4} and
\oiii\ lines.  Since \ion{C}{4} is a doublet, it is tricky to perform
a direct comparison of the two lines \citep{hainlineetal2011}. As
throughout the paper, we assume that the \oiii\ provides a model for
the narrow-line region.  We then assume the same redshift for both the
\oiii\ and \ion{C}{4} lines, and construct a model \ion{C}{4}
profile as the superposition of two lines with the same line shape in
velocity space as the \oiii\ line. We convert the FIRE spectra to
vacuum wavelengths, place the two doublet components at their
laboratory wavelengths, and impose a line ratio of 1:2 \citep{hamannetal1995} 
(although we also tried 0.9:1 as above; the results do not change).

\vbox{ 
\vskip +5mm
\hskip -1mm
\includegraphics[scale=0.55,angle=0]{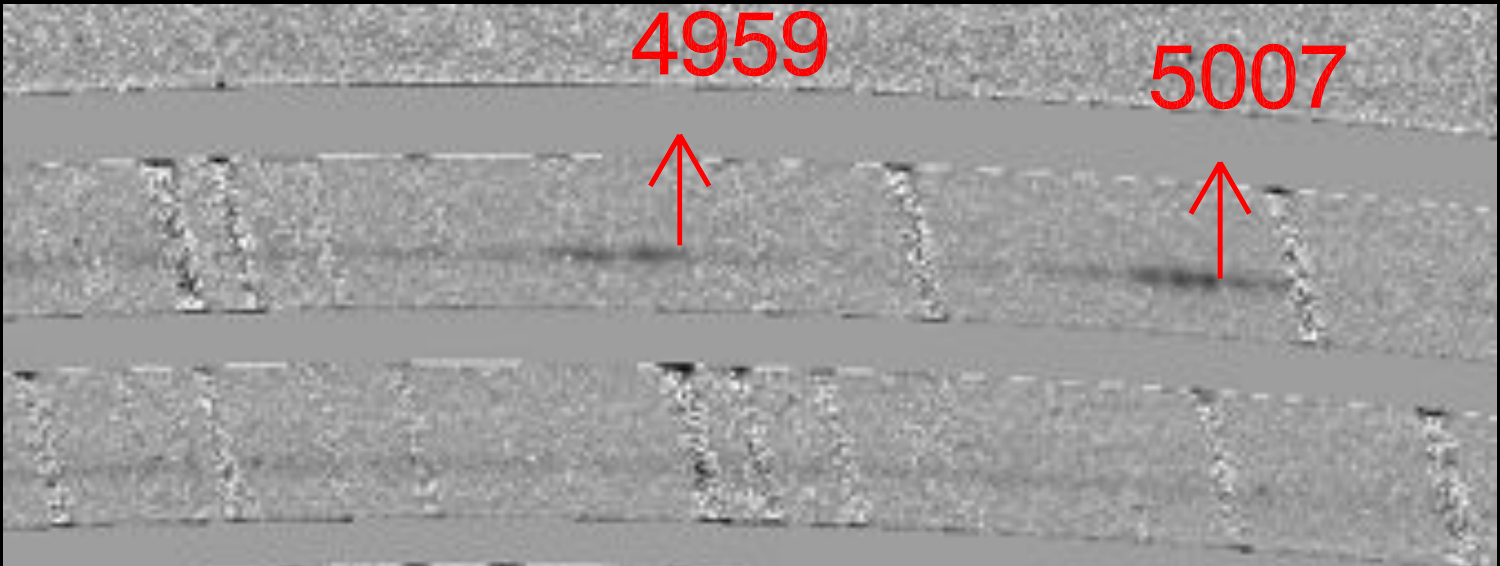}
}
\vskip -0mm
\figcaption[]{
Flat-fielded and sky-subtracted two-dimensional 
spectrum for the quasar SDSS J1339$+0441$, which has multiple clumps in the 
[O{\tiny III}] line shown here.  The grey bands represent gaps between orders, 
and the galaxy continuum is clearly visible in both orders shown here. 
The arrows roughly indicate zero 
velocity for the 4959 (left) and 5007 (right) lines. The clumps are 
$\sim 300$~\kms\ apart, are far from strong sky lines, and 
are most likely intrinsic.
\label{fig:twodclump}}
\vskip 5mm

Much like all other targets studied in this paper, the \ion{C}{4} line
in both triple-peaked objects has a broad base that cannot be
explained by a simple superposition of the \oiii\ lines (Figure
\ref{fig:onedclump}).  Furthermore, some narrow absorption must be
superposed on the \ion{C}{4} emission profiles to fully explain their
velocity structure.  We indicate in dotted red lines the locations of
these putative absorbers in each system.  However, this additional
absorption is only marginally broader than one pixel, making it quite
suspect, and the doublet line ratios are not sensible.

In summary, these multi-peaked objects have become more intriguing and
more mysterious with our additional NIR data.  On the one hand, in 
one case we found velocity structure in the \oiii\ line, which,
unlike in the \ion{C}{4}, cannot be due to blue-shifted absorption.
On the other hand, we cannot fully explain the observed \ion{C}{4}
kinematics with a simple superposition of the \oiii\ line profiles.
We either must postulate absorption in addition to real velocity
structure, or accept that the line shapes are unrelated.  For
instance, perhaps the \ion{C}{4} arises from the broad-line region, or an
intermediate-line region sitting between the classic narrow and
broad-line regions \citep{brothertonetal1994}.

\section{Discussion and Summary}
\label{sec:Discussion}

We started with a sample of Type II quasar candidates selected from
BOSS, with relatively narrow and high-EW \ion{C}{4} emission lines.
In this paper we present NIR spectroscopy in the $JHK$ bands, covering
\hbeta\ and \oiii\ emission in all cases, and \halpha\ in most.  Our
primary motivation is to investigate the nature of these targets,
specifically whether they are in fact obscured quasars.  Overall, our
analysis points to a population of moderately obscured ($A_B \sim 0-3$
mag) quasars powered by $\sim
10^9$~\msun\ BHs radiating at a healthy $\sim 10\%$ of their Eddington
limits.

From Paper I, we know that our quasar targets have UV luminosities
that are too high to be described by star formation alone.  On the
other hand, we did detect polarization at the level of $\sim 3\%$ in
the two observed targets (Paper I), larger than the $0.5\%$ typical of
unobscured quasars \citep{berrimanetal1990}. Furthermore, we see 
evidence that the UV luminosities are somewhat suppressed, in the
broad-band SED of SDSS J0958+0135 presented in Paper I and also in the
low ratio of UV to \emph{WISE} luminosity (Figure \ref{fig:wiselum}).  
Thus, the continuum measurements support a moderate-obscuration
scenario.

We use multiple techniques to estimate the level of extinction. 
The narrow-line equivalent widths point to modest extinction; in
general \oiii\ has higher EW than a typical unobscured quasar, but not
as high as the locus of Type II quasars at $z \approx 0.5$
\citep{zakamskaetal2003}.  Turning the observed EW discrepancy into an
estimate of the extinction produces $0 < A_B < 3$ mag 
or $A_V \approx 0-2.2$ mag.  Based on 
this $A_B$, the intrinsic UV luminosities would be boosted by 
$0.5 -1 $ dex.  

We also estimate the extinction using the non-detection of broad
\hbeta, and find rough agreement with the reddening estimates
based on \oiii\ EW. The broad permitted line widths are also
consistent with a moderate-obscuration scenario.  Specifically, we detect
significant broad emission lines in \halpha\ with line widths ranging
from $1000$ to $7500$~\kms.  In the UV, on the other hand, the
\ion{C}{4} line fluxes are fainter than expected based on \halpha\ 
(assuming an SMC extinction law), and show
considerably narrower widths, ranging only from $1000$ to $4500$~\kms. 
Of course, by originally selecting narrow \ion{C}{4} lines, we naturally identified
targets at one extreme end of the line-width ratio distribution. 

At least part of the broad emission lines in the UV are likely
unobservable due to extinction.  Whether the broadest components are
preferentially obscured, leading to the narrow line widths, or whether
we simply lack the S/N to find the extended (faint) broad wings
is difficult to determine from these data.  However, at least part of the
broad-line region is directly transmitted in the UV.  This broad
component in \ion{C}{4} also explains the low ratios of \ion{He}{2} to
\ion{C}{4} reported in Paper I, which are consistent with
broad-line rather than narrow-line objects
\citep[e.g.,][]{nagaoetal2006}.

The targets presented here display broad \halpha, and thus do not obey
classic definitions of Type~II active nuclei
\citep[e.g.,][]{khachikyanweedman1971}.  Furthermore, in general, the
extinction associated with classical, optically selected Type~II
sources is larger than measured here.  For lower luminosity Seyfert
galaxies, $A_{\rm V}$ has been estimated using the ratios of
Pa$\alpha$ to \halpha, with values of more than 10 being common
\citep[e.g., ][]{rixetal1990,goodrichetal1994,veilleuxetal1997}.  In
Type II quasars at $z \approx 0.5$, the weak continuum and lack of
strong NIR emission suggests $A_{\rm B}> 13$
\citep{zakamskaetal2005}. At more comparable redshifts to our targets,
composite spectral energy distribution modeling of Type II AGN with a
galaxy+AGN template yields similarly high values of $A_B > 13$ in most
cases
\citep[e.g.,][]{hickoxetal2007,mainierietal2011,hainlineetal2012,
  assefetal2013,lussoetal2013}.  The obscuration levels reported here,
in contrast, are comparable to those seen in so-called ``Seyfert 1.8''
galaxies \citep[e.g.,][]{osterbrock1981,rixetal1990}, and the large
ratios of narrow-to-broad line FWHM seen in \halpha\ are also
characteristic of this moderately obscured class.

A more luminous population with interestingly similar characteristics
are ``red quasars,'' selected as Type~I reddened quasars in the
rest-frame optical/NIR
\citep[e.g.,][]{glikmanetal2007,urrutiaetal2009,urrutiaetal2012,banerjietal2012,
  glikmanetal2013,banerjietal2013}.  When selected in the rest-frame
optical/NIR, such objects make up $\sim 20\%$ of the quasar population
\citep[][]{glikmanetal2012,elitzur2012}.  Like our targets, these 
so-called red quasars have moderate extinction, $A_V \approx 0.3-4.5$ mag,
  such that broad \halpha\ is observed. Our targets are typically more
  luminous in \emph{WISE} by at least an order of magnitude than those 
in \citet{glikmanetal2012}, but are
  also typically found at higher redshifts.  The Banerji objects are
  selected from deeper NIR photometry and thus comprise somewhat more
  extincted ($A_V \approx 2-6$ mag) and more intrinsically luminous
  objects (by factors of a few) at comparable redshifts to ours.
Eventually we will be able to compare these samples in the rest-frame
UV as well (Glikman, private communication).  Sources selected in
  the mid-IR \citep[e.g.,][]{deyetal2008,sternetal2012,assefetal2013}
  tend to be factors of a few more luminous in the MIR and redder than
  our rest-frame UV selected sample \citep[e.g.,][]{wujetal2012}.

A few quasars with similarly moderate extinction inferred from the
rest-frame optical/UV have also been uncovered in X-ray surveys,
sometimes with significant obscuring columns in the X-ray
\citep[e.g.,][]{almainietal1995,georgantopoulosetal1999,georgantopoulosetal2003,
  piconcellietal2005}.  X-ray spectroscopy of some of our quasars
would allow the relative levels of X-ray and optical/UV obscuration to
be assessed. Substantial apparent differences between such obscuration
levels are often found, likely owing to the presence of some X-ray
absorption within the dust-sublimation radius
\citep[e.g.,][]{maiolinoetal2001}.

Obviously, each selection method imposes a preferred distribution of
reddening and luminosity on the final sample.  \citet{assefetal2013}
have recently measured a distribution of $E(B-V)$ for luminous
\emph{WISE}-selected quasars at $z < 1$.  Interestingly, they find two
natural breaks in the distribution.  The first is at $E(B-V) \approx
0.15, A_V \approx 0.5$ mag, corresponding roughly to the sources
observed here.  They find a second transition at $E(B-V) \approx 2,
A_V \approx 6$ mag. They postulate that the moderate extinction may
arise on galaxy-wide scales while the second transition is due to the
orientation of a dusty torus in the nucleus.  Perhaps our \hst\
imaging will allow us to distinguish these two scenarios for our
candidates (Strauss et al. in prep.).

Taken altogether, our observations paint a nuanced picture of a
mildly-obscured population of high-redshift quasars selected with the
BOSS survey.  How they fit into the broader picture of quasar
demographics at this epoch remains to be seen.  As a class of objects
with obscuration intermediate between Type~I and Type~II sources, they
may be at intermediate orientation in a unification picture, or just
result from a patchy torus
\citep[e.g.,][]{elitzur2012}. Alternatively, the host galaxy may
provide some large fraction of the obscuration, in an evolutionary
picture \citep[e.g.,][]{sandersetal1988,lacyetal2007,hopkinsetal2006}.
We may hope to get some clues to 
their star formation properties from higher S/N NIR spectra.
We also have many more BOSS targets to study; the sample properties of the
targets presented in Paper I are heterogeneous, and the faintest
targets may yet harbor some bona-fide Type II quasars. We are also
planning to obtain uniform NIR imaging, to preferentially select the
faintest targets in the rest frame optical, which are most likely to
be heavily obscured.  Combining broad-band SED fitting, \hst\ imaging,
spectropolarimetry, and more NIR spectroscopy, may provide a more
complete picture of the nature of these enigmatic targets.

\acknowledgements 
We thank the referee for a very insightful report that improved the
quality of this manuscript.  We thank E. Glikman, G. Richards,
R. Riffel, and G. Zhu for useful discussions, and gratefully thank the
Gemini, Magellan, and APO staff who helped with both Phase II
preparations and observations on the mountain. MAS and RA thank the
support of NSF grant AST-0707266.  Funding for SDSS-III has been
provided by the Alfred P. Sloan Foundation, the Participating
Institutions, the National Science Foundation, and the U.S. Department
of Energy Office of Science. The SDSS-III web site is
http://www.sdss3.org/.

SDSS-III is managed by the Astrophysical Research Consortium for the
Participating Institutions of the SDSS-III Collaboration including the
University of Arizona, the Brazilian Participation Group, Brookhaven
National Laboratory, Carnegie Mellon University, University of
Florida, the French Participation Group, the German Participation
Group, Harvard University, the Instituto de Astrofisica de Canarias,
the Michigan State/Notre Dame/JINA Participation Group, Johns Hopkins
University, Lawrence Berkeley National Laboratory, Max Planck
Institute for Astrophysics, Max Planck Institute for Extraterrestrial
Physics, New Mexico State University, New York University, Ohio State
University, Pennsylvania State University, University of Portsmouth,
Princeton University, the Spanish Participation Group, University of
Tokyo, University of Utah, Vanderbilt University, University of
Virginia, University of Washington, and Yale University.


\end{document}